\documentclass[12pt]{article}
\usepackage{epsfig,amssymb,euscript,bbold}
\usepackage{amsfonts}
\usepackage{eufrak}
\usepackage{mathrsfs}
\usepackage{graphicx}
\usepackage{epsfig}
\usepackage{bm}
\usepackage{amsmath}
\def\be#1\ee{\begin{equation}#1\end{equation}}
\newcommand{\bea}{\begin{eqnarray}}
\newcommand{\eea}{\end{eqnarray}}
\newcommand{\ba}{\begin{array}}

\newcommand{\ea}{\end{array}}

\def\bbox{{\,\lower0.9pt\vbox{\hrule \hbox{\vrule height 0.2 cm
\hskip 0.2 cm \vrule height 0.2 cm}\hrule}\,}}
\newcommand{\dsl}{\pa \kern-0.5em /}
\renewcommand{\t}{\theta}

\newcommand{\G}{\Gamma}

\newcommand{\nn}{\nonumber \\}

\def\w{\wedge}
\def\mbb{\mathbb{R}}

\def\l{\lambda}
\def\s{\sigma}
\def\r{\rho}
\def\a{\alpha}
\def\e{\epsilon}
\def\d{\delta}

\def\ds{\raise.15ex\hbox{/}\kern-.57em\partial}
\def\Ds{\,\raise.15ex\hbox{/}\mkern-13.5mu D}
\newcommand{\dd}{\mathrm{d}}
\newcommand{\DD}{\mathrm{D}}
\newcommand{\ee}{\mathrm{e}}

\newcommand{\rep}[1]{\mathbf{#1}}

\newcommand{\pr}{\partial_{\hat{\r}}}
\newcommand{\dn}{\dd_7}

\begin{document}

\makeatletter
\renewcommand{\theequation}{\thesection.\arabic{equation}}
\@addtoreset{equation}{section}
\makeatother

\baselineskip 18pt

\begin{titlepage}

\vfill

\begin{flushright}
Imperial/TP/2007/OC/01\\
\end{flushright}

\vfill

\begin{center}
   \baselineskip=16pt
   {\Large\bf Global geometry of the supersymmetric $AdS_3/CFT_2$
     correspondence in M-theory}
   \vskip 2cm
      Pau Figueras$^{a}$, Ois\'{\i}n A. P. Mac Conamhna$^{b}$ and Eoin \'{O} Colg\'{a}in$^{b}$
   \vskip .6cm
      \begin{small}
      \textit{$^{a}$Departament de F{\'\i}sica Fonamental}\\
	\textit{Universitat de
	Barcelona, \\Diagonal 647, E-08028 Barcelona,
	Spain}\\
      \medskip
      \textit{$^{b}$Blackett Laboratory, Imperial College\\
        London, SW7 2AZ, U.K;\\The Institute for Mathematical
        Sciences,\\Imperial College, London SW7 2PG, UK.}
        \end{small}
   \end{center}

\vskip .3 in
\begin{center}
{\texttt{pfigueras@ffn.ub.es, o.macconamhna@imperial.ac.uk, eoin.ocolgain@imperial.ac.uk}}
\end{center}

\begin{center}
\textbf{Abstract}
\end{center}
We study the global geometry of a general class of spacetimes of
relevance to the supersymmetric $AdS_3/CFT_2$ correspondence in
eleven-dimensional supergravity. Specifically, we study spacetimes
admitting a globally-defined 
$\mathbb{R}^{1,1}$ frame, a globally-defined frame bundle with structure group
contained in $Spin(7)$, and an $AdS_3$ event horizon or conformal boundary. We show how the
global frame bundle may  
be canonically realised by globally-defined null sections of the spin
bundle, which we use to truncate eleven-dimensional supergravity
to a gravitational theory of a frame with structure group
$Spin(7)$, $SU(4)$ or $Sp(2)$. By imposing an $AdS_3$ boundary
condition on the  
truncated supergravity equations, we define the geometry of all
$AdS_3$ horizons or boundaries which can be obtained from solutions of
these truncations. In the 
most generic case we study, we reproduce the most general conditions
for an $AdS_3$ manifold in M-theory to admit a Killing spinor. As a
consistency check on our definitions of $AdS$ geometries we
verify that they are satisfied by known gauged supergravity $AdS_3$
solutions. We discuss future applications of our results. 

\begin{quote}

\end{quote}

\vfill

\end{titlepage}

\setcounter{equation}{0}

\section{Introduction}
The formulation of the AdS/CFT correspondence \cite{malda} has
stimulated intense and ongoing interest in the geometry of Anti-de
Sitter manifolds, and their quantum field theoretic description, in
string and M-theory. By now there exists an extensive list of explicit
supersymmetric $AdS$ solutions of the field equations of
ten- and eleven-dimensional supergravity, and an extensive range of
solution generating techniques - for example, by taking the
near-horizon limit of an elementary or wrapped brane configuration
\cite{25}-\cite{47}, or by applying the gravity dual of a marginal
field theory deformation to a known solution \cite{malda2}, \cite{beta}. More
generally, there are many known Minkowski solutions which asymptote to
$AdS$, either at a horizon or a conformal boundary. The elementary
brane solutions describe interpolations from a conical special holonomy
manifold at a spacelike infinity to an internal $AdS$ spacelike infinity
associated to an event horizon; and there are many known globally
Minkowski and asymptotically $AdS$ solutions admitting an
interpretation as the dual of an RG flow to a superconformal fixed
point, for example \cite{nick1}-\cite{nick3}. More generally still, there are
Minkowski solutions without an $AdS$ region which may be interpreted
as dual to confining gauge theories, such as the warped deformed
conifold \cite{ks}. 

Our primary goal in this paper is to define the general global
features of the geometry of supersymmetric spacetimes in eleven
dimensions which are globally or locally $AdS_3$. The globally $AdS_3$
spacetimes arise as the horizon manifolds of branes, or the fixed
point manifolds of RG flows; the locally $AdS_3$ spacetimes can be
interpreted as the full brane or RG flow solutions. Our approach is a
direct continuation of that of \cite{wrap}, \cite{eoin}. For our basic 
set-up, we require the global 
existence of a warped $\mbb^{1,1}$ frame, with a global reduction of
the frame bundle on the transverse space; the metric is given by
\bea\label{1.4}
\dd s^2=2e^+\otimes e^-+\dd s^2(\mathcal{M}_8)+e^9\otimes e^9,
\eea
where we impose that $e^+=L^{-1}\dd x^+$, $e^-=\dd x^-$, and
$L<\infty$ globally; that $L$,
the metric on 
$\mathcal{M}_8$, and the basis one-form $e^9$ are everywhere
independent of the coordinates $x^{\pm}$; and that $e^9\neq0$
is everywhere non-vanishing. We demand that the flux respects the
Minkowski isometries; in other words, that it is given by
\bea\label{1.5}
F=e^{+-}\w H+G,
\eea
with $H$ and $G$ independent of the Minkowski coordinates,
globally. Our final assumption is that $\mathcal{M}_8$ admits a
globally-defined $G$-structure. We will study globally defined
$Spin(7)$, $SU(4)$ 
and $Sp(2)$ structures on $\mathcal{M}_8$. The existence of a globally
defined $Spin(7)$ structure on $\mathcal{M}_8$ is equivalent to the
existence of a no-where vanishing $Spin(7)$ invariant Cayley four-form $\phi$ on
$\mathcal{M}_8$. For $SU(4)$, the globally-defined forms are the
almost complex structure $J$ and the (4,0) form $\Omega$. For $Sp(2)$,
the existence of the global structure is equivalent to the existence
of a triplet of everywhere non-zero almost complex structures $J^A$,
$A=1,2,3.$ 

Our assumption of the existence of a global frame bundle is a
stronger one than the more traditional assumption of the existence of
a generic section of the spin bundle -  a globally non-vanishing
Killing spinor. All sorts of complications can potentially occur in the global
behaviour of generic sections of the spin bundle - timelike spinors becoming
null, spinors becoming parallel, and so forth - that seriously restrict
their usefulness as a global tool. Part of our motivation for 
assuming the existence of a frame bundle is that it provides
significant global control over the geometry, and these issues do not
arise. Heuristically, a second 
motivation is that the workings of AdS/CFT appear to be reflected in
the very special global properties of the relevant supergravity
solutions, and we believe that all known $AdS$, brane or RG flow
supergravity solutions satisfy this assumption. A third, more concrete
motivation for this assumption is that 
it has played an important r$\hat{\mbox{o}}$le in the recent beautiful work
on $\mathcal{N}=1$ superconformal field
theories in four dimensions and interpolations from Calabi-Yau cones
to $AdS_5\times Y^{p,q}$ manifolds in IIB \cite{kleb}-\cite{6}. The
metric and flux for these supergravity solutions are given by
\bea\label{1.1}
\dd s^2&=&\left[1+\frac{1}{R^4}\right]^{-1/2}\dd s^2
(\mbb^{1,3})+\left[1+\frac{1}{R^4}\right]^{1/2}[\dd R^2+R^2\dd
s^2(\mathcal{M}_5)],\nn
F&=&(1+\star)\mbox{Vol}_{\mbb^{1,4}}\w\dd \left[1+\frac{1}{R^4}\right]^{-1},
\eea
where $\dd s^2(\mathcal{M}_5)$ is a Sasaki-Einstein metric on $Y^{p,q}$. As
$R\rightarrow\infty$, the metric asymptotes to a singular Calabi-Yau cone:
\bea
\dd s^2\rightarrow \dd s^2(\mbb^{1,3})+ \dd R^2+R^2\dd
s^2(\mathcal{M}_5).
\eea
A  global geometry of this form would be singular at
$R=0$. However, in the interpolating solution, this singularity is
excised, and removed to infinity. The apex of the cone is thereby rendered non-compact, and
opens up into an internal, asymptotically $AdS_5$ region, at infinite
proper distance. The Penrose diagram, in the $t-R$ plane, for the maximal analytic
extension of this manifold \cite{gary} is shown in Figure 1. \begin{figure}[t!]
\begin{picture}(0,0)(0,0)
\put(88,93){$i^0$,}
\put(103,93){\small{$AdS_5$}}
\put(212,93){$i^0$}
\put(185,93){\small{$CY_3$},}
\put(195,50){$\mathscr{I}^-$}
\put(195,135){$\mathscr{I}^+$}
\put(120,125){$H^+$}
\put(120,59){$H^-$}
\put(150,155){$i^+$}
\put(154,26){$i^-$}
\put(137,168){$i^0$}
\put(105,168){\small{$AdS_5$},}
\put(12,168){$i^0$,}
\put(27,168){\small{$CY_3$}}
\put(30,215){$\mathscr{I^+}$}
\put(30,115){$\mathscr{I^-}$}
\put(77,105){$i^-$}
\put(75,230){$i^+$}

\put(154,178){$i^-$}
\put(150,305){$i^+$}
\put(195,200){$\mathscr{I}^-$}
\put(195,285){$\mathscr{I}^+$}
\put(88,243){$i^0$,}
\put(103,243){\small{$AdS_5$}}
\put(212,243){$i^0$}
\put(185,243){\small{$CY_3$},}
\put(120,275){$H^+$}
\put(120,209){$H^-$}
\end{picture}
\centering{\includegraphics[height=12cm]{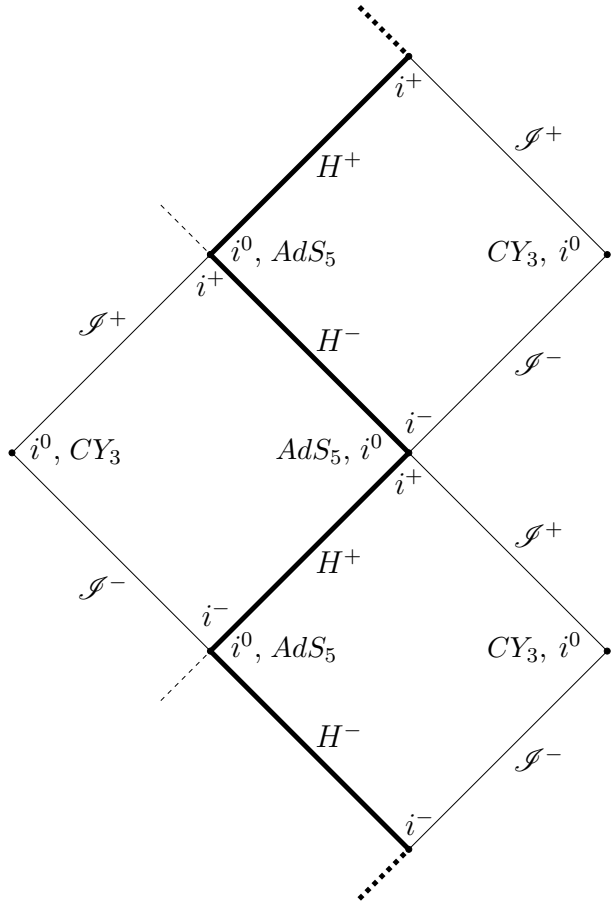}}
\caption{Penrose diagram for the maximal analytic extension of an
      interpolation from a Calabi-Yau cone to $AdS_5$ in IIB.}
\label{fig:CPCY3}
\end{figure}
An important global assumption in identifying
the geometric dual of a-maximisation \cite{6} is that the Calabi-Yau
singularity is Gorenstein. This means that the incomplete special
holonomy manifold obtained upon excising the singularity is globally
Calabi-Yau; it admits an everywhere non-vanishing complex structure
and holomorphic three-form. An equivalent statement of this assumption
is that the interpolating solution (where the singularity is indeed
excised, and removed to infinity) admits a global reduction of the
frame bundle to a principal $SU(3)$ sub-bundle, on an incomplete region of
spacetime bounded by the special holonomy asymptotics and the $AdS$
horizon - a causal diamond of the Penrose diagram. Analytic
extension of the frame bundle across an event horizon appears to be
facilitated by the doubling of supersymmetry on the $AdS$ horizon
manifold. However we do not explore the issue of analytic extension
across a horizon any further here, and we
henceforth restrict attention to regions of spacetime bounded by 
asymptopia and $AdS$ horizons, admitting a global reduction of the
frame bundle. This restriction to a causal
diamond of a Penrose diagram is in any event in keeping with more general
ideas about holography, and also plays an important r$\hat{\mbox{o}}$le in the
quantum gravity of de Sitter space \cite{strominger}, \cite{susskind}. 

We will now begin to explore what information about the geometry of
\eqref{1.4}, \eqref{1.5} we can
extract, from eleven-dimensional 
supergravity, given our global assumptions. Eleven
dimensional supergravity is not designed to 
manipulate frame bundles directly - the Killing spinor equation is
instead an equation
for sections of the spin bundle. In demanding the
existence of a globally defined frame bundle, we have not
assumed any a priori realisation of the frame bundle by sections
of the spin bundle. Therefore, in order to use eleven-dimensional
supergravity, we must find a way of
associating globally defined sections of the spin bundle to a globally-defined
frame bundle. By this we mean finding the Killing spinors whose bilinears
produce the structure forms. Clearly, they should be singlets of the structure
group. They may be selected in a natural way, by using the
Clifford action of the structure forms on the eleven dimensional spin
bundle. This Clifford action is defined for an $n$-form $A$ on
$\mathcal{M}_8$ by
\bea
A\cdot\eta=\frac{1}{n!}A_{i_1...i_n}\G^{i_1...i_n}\eta,
\eea
where $\eta$ is a Majorana spinor in eleven dimensions and the $\G^i$
are eleven-dimensional gamma-matrices, with $i=1,...,8$. Taking the
example of $Spin(7)$, the Clifford action $\phi\cdot\eta$ decomposes
an arbitrary spinor $\eta$ into modules of the structure group of
$\phi$; each module is an eigenmodule of the Clifford action of the
structure form, with a different eigenvalue. What distinguishes the
singlets, in general, is that they are the modules of highest norm
eigenvalue. Thus,
normalising appropriately, we may obtain the singlets of the structure
group $Spin(7)$ as solutions of
\bea\label{gfd}
\frac{1}{14}\phi\cdot\eta=-\eta.
\eea  
Any globally-defined Killing spinors which give a
realisation of the globally defined $Spin(7)$ structure must lie in
the two-dimensional solution space of \eqref{gfd}. We will
impose one further condition on all Killing spinors throughout this
paper: we demand that all Killing spinors have a definite
$\mbb^{1,1}$ chirality; in other words, that they are eigenspinors of
$\G^{+-}$. Combining this restriction with \eqref{gfd}, we get the
global definition of the Killing spinors realising a $Spin(7)$
structure:
\bea\label{fds}
\G^{+-}\frac{1}{14}\phi\cdot\eta=\pm\eta.
\eea
On a special holonomy manifold, in less geometrical language, this
would be called the kappa-symmetry projection for a probe M5 brane
wrapped on a Cayley four-cycle. 

For our global $Spin(7)$ structure, there are two distinct Killing
spinor realisations; one 
where only one solution of \eqref{fds} exists globally, and one where
both solutions exist globally. When given a wrapped-brane
interpretation, the first case can be associated to interpolating
solutions involving deformations of the normal bundle of a Cayley 
four-cycle of a $Spin(7)$ manifold. The second case is associated to
interpolations from a 
$Spin(7)$ cone to an $AdS_4$ horizon (foliated by $AdS_3$ leaves). For $SU(4)$
and $Sp(2)$, finding the spinorial realisations of the 
global frame bundle is very similar, and is discussed in detail in
section 3.

Having found the spinorial realisation of the frame bundle, we may truncate eleven-dimensional supergravity, 
globally, to a gravitational theory in eleven dimensions for a frame
bundle which is not $Spin(1,10)$, but rather $Spin(7)$, in the generic
case we consider. One
may re-interpret the BPS conditions for the globally-defined 
Killing spinor(s) realising the $Spin(7)$-structure - together with
such components of the field equations and Bianchi identity as are
not implied by their existence - as being instead the truncation of the
field equations of eleven-dimensional supergravity to a
frame bundle with structure group $Spin(7)$; in effect, a classical theory
of $Spin(7)$ gravity in eleven dimensions. 

Let us illustrate this truncation for the most generic case we
consider in this paper; a global $Spin(7)$ structure realised by a single
solution of \eqref{fds}. We will refer to this as a Cayley structure,
a Cayley frame bundle, or simply Cayley geometry,
henceforth. All other cases we study may be regarded as
particular cases of this one, with more restrictive global
conditions. The BPS conditions with our frame and a single globally
defined Killing solution of \eqref{fds} may easily be obtained from the results of
\cite{j}. The conditions on the intrinsic torsion of the globally-defined
$Spin(7)$-structure are
\bea\label{poi}
e^9\w\left[-L^3e^9\lrcorner\; \dd
  (L^{-3}e^9)+\frac{1}{2}\phi\lrcorner\;\dd\phi\right] &=&0,\\\label{poiu}
(e^9\w+\star_9)[e^9\lrcorner\;\dd (L^{-1}\phi)]&=&0.
\eea  
Here $\star_9$ denotes the Hodge dual on the space transverse to the
Minkowski factor. The operation $\lrcorner$ is defined, for an
$n$-form $A$ and an $m$-form $B$, $m>n$, by
\bea
A\lrcorner
B_{\mu_{n+1}...\mu_m}=\frac{1}{n!}A^{\mu_1...\mu_n}B_{\mu_1...\mu_n\mu_{n+1}...\mu_m}.
\eea
Then the flux is given by
\bea
F=\dd (e^{+-9})-\star\dd(e^{+-}\w\phi)-\frac{L^{10/7}}{2}e^9\lrcorner\;\dd(L^{-10/7}\phi)+\frac{1}{4}\phi\diamond[e^9\lrcorner\;(e^9\w\dd
e^9)]+F^{\rep{27}}.
\eea
We have defined the operation $\diamond$
for an $n$-form $A$ and a two-form $B$ on $\mathcal{M}_8$ according to
\bea
A\diamond B=nA_{[i_1...i_{n-1}}^{\;\;\;\;\;\;\;\;\;\;i_m}B_{i_n]i_m}.
\eea
Observe that $\phi\;\diamond$ is a map
$\phi\;\diamond:\Lambda^2(\mathcal{M}_8)\rightarrow\Lambda^4_{\mathbf{7}}(\mathcal{M}_8)$. The
$F^{\rep{27}}$ term in the flux is a four-form on $\mathcal{M}_8$ in the
$\rep{27}$ of $Spin(7)$ which is unfixed by the truncation. Thus, the general
equations for the truncation of eleven dimensional supergravity to 
Cayley geometry are the torsion conditions \eqref{poi}
and \eqref{poiu}, coupled to the Bianchi identity
\bea
\dd F=0
\eea
and, as it turns out (all other field equation components being
implied), the $+-9$ component of the four-form field
equation\footnote{This equation can of course receive quantum corrections, as
  can the Killing spinor and Einstein equations, but we will ignore them.} 
\bea
\star\left(\dd \star F+\frac{1}{2}F\w F\right)=0.
\eea 

Having obtained the truncated supergravity equations, we must also
specify the boundary conditions of interest to us. We will impose the
existence of an $AdS_3$ region, which we view as being associated
either to a horizon or a conformal boundary of a globally Minkowski
solution. It will be very interesting to explore more sophisticated
boundary conditions in the future. Because of the global structure,
topological considerations will be important in doing
this. Generically, one would expect a solution with an $AdS_3$ region
to go to some flux geometry at other asymptopia. But one could easily
imagine imposing more 
specialised boundary conditions, such as the existence of more than
one $AdS$ region - as relevant for the dual of an RG flow between
fixed points. From a mathematical point of view, perhaps the most
interesting additional boundary condition would be
asymptotic fall-off of the flux. This is because far from a
gravitating source, the spacelike asymptotics necessarily, and
automatically, have $Spin(7)$ holonomy; they must be Ricci-flat by
Einstein, and special holonomy by the frame bundle. Solutions of the truncated
supergravity equations with these boundary conditions describe
interpolations from special holonomy spacelike asymptopia to $AdS$
horizons. Because of the global structure, the $AdS$
horizon geometry of an interpolating solution will be intimately
related to that of the asymptotic special holonomy manifold. We will
return to a discussion of these boundary conditions in the conclusions. 

In this paper, we will impose the most
general $AdS_3$ boundary condition on the supergravity
truncations we study, leaving additional specialisations for the future. As we shall explain in detail, we do this by inserting the most general
locally $AdS_3$ frame into the globally-defined
$\mathbb{R}^{1,1}$ frame, and converting the equations for the
globally-defined Minkowski structure into a set of equations for the
locally-defined $AdS_3$ structure. For the generic case of Cayley geometry,
the local $AdS$ structure is $G_2$, with associative three-form $\Phi$ and
co-associative four-form $\Upsilon$. We shall see that locally, the
metric may always be cast in the form
\bea
\dd s^2=\frac{1}{\l m^2}\left[\dd s^2(AdS_3)+\frac{\l^3}{4\sin
    ^2\theta}\dd\r\otimes\dd\r\right]+\dd s^2(\mathcal{N}_7),
\eea
where the $G_2$ structure is defined on $\mathcal{N}_7$, and $\l$,
$\theta$ and the frame on $\mathcal{N}_7$ are independent of the
$AdS_3$ coordinates. The
restrictions on the intrinsic torsion of the locally defined $G_2$
structure may be expressed as
\bea
\hat{\r}\w\dd(\l^{-1}\Upsilon)&=&0,\\
\l^{5/2}\dd\left(\l^{-5/2}\sin\theta\mbox{Vol}_7\right)&=&-4m\l^{1/2}\cos\theta\hat{\r}\w\mbox{Vol}_7,\\
\dd\Phi\w\Phi&=&\frac{4m\l^{1/2}}{\sin\theta}(4-\sin^2\theta)\mbox{Vol}_7-2\cos\theta\star_8\dd\log\left(\frac{\l^{3/2}\cos\theta}{\sin^2\theta}\right);\nn
\eea
the flux is given by
\bea
F&=&\frac{1}{m^2}\mbox{Vol}_{AdS_3}\w\dd[\rho-\l^{-3/2}\cos\theta]\nn&+&
\frac{\l^{3/2}}{\sin^2\theta}\Big(\cos\theta+\star_8\Big)\Big(\dd[\l^{-3/2}\sin\theta\Phi]-4m\l^{-1}\Upsilon\Big)+2m\l^{1/2}\Phi\w\hat{\r},
\eea
and the definitions of $\star_8$ and the basis one-form
$\hat{\r}$ hopefully are obvious\footnote{The orientations for the various Hodge
  stars will be specified when they next appear.}. 
In \cite{J+D}, Martelli and Sparks gave a classification
of all minimally supersymmetric $AdS_3$ spacetimes in M-theory; the
conditions we have obtained on the local $G_2$ structure of an $AdS_3$
region in Cayley geometry are identical to
theirs\footnote{Up to a minor discrepancy in (3.14) of \cite{J+D} which we have
  corrected.}. We regard these conditions as being valid locally on
the horizon or conformal boundary of a globally Minkowski solution, or
globally for a globally $AdS$ solution of Cayley geometry\footnote{A
  subtlety in the global validity of these conditions for globally
  $AdS$ manifolds is discussed in section 5.}.

As we have said, we study truncations of eleven-dimensional
supergravity to several different frame
bundles, with different spinorial realisations. For
a $Spin(7)$ bundle, we study spinorial realisations by either one or
two globally defined Killing spinors. We refer to the resulting
truncations of eleven-dimensional supergravity as Cayley or $Spin(7)$
geometry, respectively. The $AdS_3$ conditions we derive from Cayley
geometry define the geometry of all M-theory duals of $N=(1,0)$
two-dimensional CFTs. The $AdS_3$ conditions we derive from $Spin(7)$
geometry reproduce the $AdS_4\times$Weak $G_2$ Freund-Rubin solutions,
with the $AdS_4$ foliated by $AdS_3$ leaves. For an $SU(4)$ frame
bundle we study three distinct spinorial realisations. The first, with
the maximal number (four) of globally-defined Killing spinors,
produces a truncation we refer to as $SU(4)$ geometry. The $AdS_3$
conditions we derive from $SU(4)$ geometry produce the
Freund-Rubin $AdS_4\times SE_7$ solutions. The other two spinorial
realisations of an $SU(4)$ frame bundle we study have two globally defined
Killing spinors. We refer to the associated truncations as
K\"{a}hler-4 and Special Lagrangian-4 
(SLAG) geometry. Given a wrapped brane interpretation, one would say
that a solution of these truncations described an M5-brane wrapped on,
respectively, a K\"{a}hler or SLAG four-cycle of a Calabi-Yau
four-fold, with a membrane extended in the directions transverse to
the Calabi-Yau and intersecting the fivebrane in a string. We believe
that the $AdS_3$ conditions we derive from 
SLAG geometry define all M-theory duals of $N=(1,1)$ CFTs; and
similarly that the $AdS_3$ conditions we derive from K\"{a}hler-4
geometry (together with the $AdS_3$ conditions of \cite{wrap} from
co-associative geometry with a global Minkowski $G_2$ frame bundle)
define all M-theory duals of $N=(2,0)$ CFTs. For an $Sp(2)$ frame
bundle, we again study three distinct spinorial realisations. The first, with
the maximal number (six) of globally-defined Killing spinors,
produces a truncation we call $Sp(2)$ geometry. Again, the $AdS_3$
conditions we derive from $Sp(2)$ geometry just give the
appropriate Freund-Rubin solutions, this time the direct product of
$AdS_4$ with a Tri-Sasaki-Einstein manifold. The other two spinorial
realisations of an $Sp(2)$ frame bundle we study have three globally defined
Killing spinors. We refer to the associated truncations of eleven
dimensional supergravity as
Quaternionic K\"{a}hler (QK) and Complex Lagrangian
(CLAG) geometry. We believe that the $AdS_3$ conditions we derive
from these truncations define all M-theory duals of $N=(3,0)$ and $N=(2,1)$
CFTs respectively.

The remainder of this paper is organised as follows. For the
convenience of the reader who is not interested in their derivation, in
section 2 we
summarise our main technical results: the truncation of eleven-dimensional
supergravity to Cayley, K\"{a}hler-4, SLAG, QK or CLAG geometry,
together with the associated  
conditions for an $AdS_3$ region. These equations are
the result of involved 
calculations. As a consistency check, we have verified that
explicit K\"{a}hler-4 and SLAG $AdS_3$ solutions, known from gauged
supergravity, satisfy our definitions of $AdS$ geometry in the
appropriate truncations, by explicitly elucidating their
structure. Since our results for K\"{a}hler-4 and SLAG geometry are
derived directly from Cayley geometry, and our results for CLAG and QK
in turn are derived from those for K\"{a}hler-4 and SLAG, this serves as a rigid
overall consistency check.      
The remainder of the paper (with the exception of the conclusions) is
concerned with the derivation of the results of section 2. In section
3, we discuss the globally-defined G-structures and spinorial
realisations thereof which are of interest to us. In section 4, we
explain in more detail how to obtain the supergravity truncation in
each case. Section 5
is concerned with the derivation of the local conditions for an
$AdS_3$ region in each truncation. Section 6 discusses the
verification of the globalised $AdS$ torsion conditions for known
solutions. Section 7 concludes with some more observations,
speculations and suggestions for future directions.

\section{Summary of results}
\label{sec:summary}
In this section, we will summarise our technical results for Cayley,
K\"{a}hler-4, SLAG, QK and CLAG geometry. In each case, we will give
the globally defined spinorial realisation of the frame bundle, the
associated truncation of eleven-dimensional supergravity, and a definition
of the geometry of an arbitrary $AdS_3$ region in the truncation. For
the K\"{a}hler-4 and SLAG geometries, as an overall consistency check,
we present a known exact solution of the $AdS$ equations,
with its structure made manifest, that we have verified satisfies our
$AdS$ conditions. 

We take positive orientation in eleven dimensions to be defined by
\bea
\mbox{Vol}_{11}=e^-\w e^+\w\frac{1}{14}\phi\w\phi\w e^9.
\eea
In every case, the globally-defined Minkowski frame
is given by \eqref{1.4} of the introduction; positive orientation for
$\star_9$, the Hodge dual on the space transverse to the Minkowski
factor, is defined by
\bea
\mbox{Vol}_9=\frac{1}{14}\phi\w\phi\w e^9.
\eea
In every case, the
truncation of eleven-dimensional supergravity to the global frame
bundle consists of the quoted torsion conditions for the
globally-defined Minkowski structure coupled to the
Bianchi identity and the $+-9$ component of the four-form field
equation. The $AdS_3$ geometries automatically solve the four-form
field equation, and for them it is in every case sufficient 
to impose the Bianchi identity in addition to the torsion conditions
to ensure that they are solutions of eleven dimensional
supergravity. For the $AdS$ geometries, the warp factor, the frame on
the transverse space, and the flux, are independent of the $AdS$
coordinates. We define the basis one-form $\hat{\r}$ in the local
$AdS$ frame in every case according to
\bea
\hat{\r}&=&\frac{\l}{2m\sin\theta}\dd\r. 
\eea
The electric flux for the $\mbb^{1,1}$ geometries, in every case,
takes the form
\bea
F_{\mbox{\small{elec}}}=\dd(e^{+-9}),
\eea
while for the $AdS$ geometries, in every case, it takes the form
\bea
F_{\mbox{\small{elec}}}=\frac{1}{m^2}\mbox{Vol}_{AdS_3}\w\dd[\r-\l^{-3/2}\cos\theta].
\eea
Now we will state our results.

\subsection{Cayley geometry}
In this case, $\mathcal{M}_8$ admits a globally-defined
$Spin(7)$ structure which is realised by a single Killing solution of
\bea
\G^{+-}\frac{1}{14}\phi\cdot\eta=-\eta.
\eea
\paragraph{Global truncation} The truncation of eleven-dimensional
supergravity to this geometry is defined by 
\bea
e^9\w\left[-L^3e^9\lrcorner\; \dd
  (L^{-3}e^9)+\frac{1}{2}\phi\lrcorner\;\dd\phi\right] &=&0,\\
(e^9\w+\star_9)[e^9\lrcorner\;\dd (L^{-1}\phi)]&=&0,
\eea  
\bea
F=\dd (e^{+-9})-\star\dd(e^{+-}\w\phi)-\frac{L^{10/7}}{2}e^9\lrcorner\;\dd(L^{-10/7}\phi)+\frac{1}{4}\phi\diamond[e^9\lrcorner\;(e^9\w\dd
e^9)]+F^{\rep{27}}.
\eea
\paragraph{AdS geometry} The geometry of an $AdS_3$ region in this truncation is as
follows. Locally the metric may be cast in the form 
\bea
\dd s^2=\frac{1}{\l m^2}\left[\dd s^2(AdS_3)+\frac{\l^3}{4\sin
    ^2\theta}\dd\r\otimes\dd\r\right]+\dd s^2(\mathcal{N}_7),
\eea
where $\mathcal{N}_7$ admits a $G_2$ structure, with associative
three-form $\Phi$ and co-associative four-form $\Upsilon$. The torsion
conditions are
\bea
\hat{\r}\w\dd(\l^{-1}\Upsilon)&=&0,\\
\l^{5/2}\dd\left(\l^{-5/2}\sin\theta\mbox{Vol}_7\right)&=&-4m\l^{1/2}\cos\theta\hat{\r}\w\mbox{Vol}_7,\\
\dd\Phi\w\Phi&=&\frac{4m\l^{1/2}}{\sin\theta}(4-\sin^2\theta)\mbox{Vol}_7-2\cos\theta\star_8\dd\log\left(\frac{\l^{3/2}\cos\theta}{\sin^2\theta}\right);\nn
\eea
and the magnetic flux is given by
\bea
F_{\mbox{\small{mag}}}&=&
\frac{\l^{3/2}}{\sin^2\theta}\Big(\cos\theta+\star_8\Big)\Big(\dd[\l^{-3/2}\sin\theta\Phi]-4m\l^{-1}\Upsilon\Big)+2m\l^{1/2}\Phi\w\hat{\r}.
\eea
Positive orientation on the space transverse to the $AdS$ factor is
defined by $\frac{1}{7}\Phi\w\Upsilon\w\hat{\r}$.

\subsection{K\"{a}hler-4 geometry}
In this case, $\mathcal{M}_8$ admits a globally-defined $SU(4)$
structure. The structure is realised by two globally defined null
Killing spinors, which are solutions of
\bea
\frac{1}{12}\G^{+-}(J\w J)\cdot\eta&=&-\eta.
\eea

\paragraph{Global truncation} The truncation of eleven-dimensional
supergravity to this geometry is defined by the torsion conditions for
the global $SU(4)$ structure 
\bea
J\lrcorner\;\dd e^9&=&0,\nn
\dd(L^{-1}\mbox{Re}\Omega)&=&0,\nn
e^9\w[J\lrcorner \;\dd J-Le^9\lrcorner\;\dd(L^{-1}e^9)]&=&0,
\eea
and the four-form 
\bea\label{2}
F&=&\dd(e^{+-9})+\frac{1}{2}\star \dd\left(e^{+-}\w J\w
J\right)+\frac{1}{4}L^2e^9\lrcorner\;\dd\left(L^{-2}J\w J\right)\nn&-&\frac{1}{4}(J\w
J)\diamond[e^9\lrcorner\;(e^9\w
\dd e^9)]+F^{\rep{20}}.
\eea  
Here $F^{\mathbf{20}}$ is a four-form on
$\mathcal{M}_8$ in the $\mathbf{20}$
of $SU(4)$ (a primitive (2,2) form) which is not fixed by the
truncation.

\paragraph{AdS geometry} The local metric for an $AdS_3$ 
region in this geometry is
\bea
\dd s^2=\frac{1}{\l m^2}\left[\dd s^2(AdS_3)+\frac{\l^3}{4\sin
    ^2\theta}\dd\r\otimes\dd\r\right]+e^7\otimes e^7+\dd s^2(\mathcal{N}_6),
\eea
where $\mathcal{N}_6$ admits an $SU(3)$ structure. Using $J$ and
$\Omega$ to denote the structure forms of this local $SU(3)$ structure
(hopefully without risk of confusion with the structure forms of the
global $SU(4)$ structure), the local $AdS_3$ torsion conditions are
\bea\label{eqn:k4jj}
\hat{\rho}\w\dd(\l^{-1}J\w J)&=&0,\\
\dd(\l^{-3/2}\sin\theta\mbox{Im}\Omega)&=&2m\l^{-1}(e^7\w\mbox{Re}\Omega-\cos\theta\hat{\r}\w\mbox{Im}\Omega),\\\label{eqn:k4de7}
J\lrcorner\;\dd
e^7&=&\frac{2m\l^{1/2}}{\sin\theta}(2-\sin^2\theta)-\cos\theta\hat{\r}\lrcorner\;\dd\log\left(\frac{\l^{3/2}\cos\theta}{\sin^2\theta}\right).
\eea
The magnetic flux is 
\bea
F_{\mbox{\small{mag}}} &=&  \frac{\l^{3/2}}{\sin^2\theta}(\cos\theta+\star_8)(\dd[\l^{-3/2}\sin\theta
J\w e^{7}
]-2m\l^{-1} J\w J  ) + 2m\l^{1/2} J\w  e^{7}\w \hat{\r}.\nn
\eea
Positive orientation on the space tranverse to the $AdS$ factor is
defined by $\frac{1}{6}J\w J\w J\w e^7\w \hat{\r}$. 

\paragraph{Exact solution} We have verified that the following is an
exact solution of the $AdS$ torsion conditions and Bianchi
identity in this truncation. Topologically, the space transverse to the $AdS$ factor is
an $S^4$ bundle over a negatively curved K\"{a}hler-Einstein
manifold. This solution was first constructed in gauged supergravity,
as the near-horizon limit of an M5 brane wrapped on a K\"{a}hler
four-cycle in a Calabi-Yau four-fold, in \cite{kim}. The metric is
given by 
\bea
\dd s^2&=&\frac{1}{\l m^2}\Big[\dd s^2(AdS_3)+\frac{3}{4}\dd
  s^2(\mbox{KE}_4)+(1-\l^3 f^2)\mbox{D}Y^a\otimes
  \mbox{D}Y^a\nn&&+\frac{\l^3}{4(1-\l^3 f^2)}\dd\r\otimes\dd\r\Big],
\eea
where
\bea
\l^3=\frac{3}{4(1+\r^2/12)},\;\; f^2=\frac{4}{9}\r^2.
\eea
 The $Y^a$, $a=1,...,4$ are constrained coordinates on an $S^3$,
 $Y^aY^a=1$. We define $K^A$, $A=1,2,3$,
 $K^AK^B=-\delta^{AB}-\e^{ABC}K^C$, to be a triplet of self-dual
 two-forms on $\mbox{KE}_4$, and we choose $K^3$ to label the
 K\"{a}hler form. We define
\bea
\mbox{D}Y^a=\dd Y^a-\frac{1}{4}K^{3cd}\omega_{cd}K^{3a}_{\;\;\;\;\;b}Y^b,
\eea
where $\omega_{ab}$ are the spin connection one-forms of
$\mbox{KE}_4$. Finally $\dd s^2(\mbox{KE}_4)$ is normalised such
that the Ricci form is given by $\mathcal{R}=-K^3$. Defining the functions
\bea 
g=\frac{\sqrt{3}}{2\l^{1/2}m},\;\; h=\frac{\sqrt{1-\l^3
    f^2}}{\l^{1/2}m},
\eea
the $SU(3)$ structure forms are given by 
\bea
e^7&=&hK^3_{ab}Y^a\mbox{D}Y^b,\nn
J&=&g^2K^3+h^2\frac{1}{2}K^3_{ab}\mbox{D}Y^a\w\mbox{D}Y^b,\nn
\mbox{Re}\Omega&=&-g^2h\left[K^2\w K^1_{ab}Y^a\mbox{D}Y^b+K^1\w K^2_{ab}Y^a\mbox{D}Y^b\right],\nn\label{eqn:K4su3}
\mbox{Im}\Omega&=&-g^2h\left[K^2\w K^2_{ab}Y^a\mbox{D}Y^b-K^1\w K^1_{ab}Y^a\mbox{D}Y^b\right].
\eea
In \cite{46}, \cite{47}, many infinite families of $AdS_3$ solutions,
generalising this one, were constructed. All these families will satisfy our
$AdS$ equations for K\"{a}hler geometry.

\subsection{Special Lagrangian geometry}
Again in this case $\mathcal{M}_8$ admits a globally defined $SU(4)$
structure. It is realised by two globally defined null Killing
solutions of 
\bea
\G^{+-}\eta&=&\pm\eta,\nn
\frac{1}{8}\G^{+-}\mbox{Re}\Omega\cdot\eta&=&-\eta.
\eea

\paragraph{Global truncation} In this case, the torsion conditions for
the global $SU(4)$ structure are 
\bea
\dd(L^{-1/2}J)&=&0,\nn
\mbox{Im}\Omega\w \dd\mbox{Re}\Omega&=&0,\nn
e^9\w[\mbox{Re}\Omega\lrcorner \;\dd\mbox{Re}\Omega-2L^{3/2}e^9\lrcorner\;\dd(L^{-3/2}e^9)]&=&0.
\eea
The flux is given by
\bea
F&=&\dd(e^{+-9})+\star \dd(e^{+-}\wedge\mbox{Re}\Omega)+\frac{1}{2}L^{7/4}e^9\lrcorner\;\dd(L^{-7/4}\mbox{Re}\Omega)\nn&-&\frac{1}{2}\mbox{Re}\Omega\diamond[e^9\lrcorner\;(e^9\w
\dd e^9)]+F^{\rep{20}}.
\eea

\paragraph{AdS geometry} The local $AdS_3$ frame and orientation are
as for K\"{a}hler-4 geometry, and again $\mathcal{N}_6$ admits an
$SU(3)$ structure. The $AdS$ torsion conditions, for the local $SU(3)$
structure forms $J$, $\Omega$, are 
\bea
e^7\w\hat{\r}\w\dd\left(\frac{\mbox{Re}\Omega}{\sin\theta}\right)&=&0, \label{eqn:slag4Re}\\
\dd(\l^{-1}\sin\theta e^7)&=&m\l^{-1/2}(J+\cos\theta e^7\w\hat{\r}),\label{eqn:slag4de7}\\
\mbox{Im}\Omega\w\dd\mbox{Im}\Omega&=&\frac{m\l^{1/2}}{\sin\theta}(6+4\cos^2\theta)\mbox{Vol}_6\w
e^7-2\cos\theta\star_8\dd\log\left(\frac{\l^{3/2}\cos\theta}{\sin^2\theta}\right).\nn
\label{eqn:slag4Im}
\eea
The magnetic flux is given by
\bea
F_{\mbox{\small{mag}}}&=&-\;\frac{\l^{3/2}}{\sin^2\theta}(\cos\theta+\star_8)(\dd[\l^{-3/2}\sin\theta\mbox{Im}\Omega]+4m\l^{-1}\mbox{Re}\Omega\w
e^7)-2m\l^{1/2}\mbox{Im}\Omega\w\hat{\r}.\nn
\eea

\paragraph{Exact solution} We have verified that the following is an
exact solution of the $AdS$ torsion conditions and Bianchi
identity in this truncation. Topologically, the eight-manifold transverse to the $AdS$
factor is an $S^4$ bundle over $\mathcal{H}^4$. Again this solution
was first constructed (in seven-dimensional gauged supergravity) in
\cite{kim}, as the near-horizon limit of an M5 brane wrapped on a SLAG
four-cycle of a Calabi-Yau four-fold.The metric is given by
\bea
\dd s^2&=&\frac{1}{\l m^2}\Big[\dd s^2(AdS_3)+\frac{2}{3}\dd
  s^2(\mathcal{H}^4)+(1-\l^3 f^2)\mbox{D}Y^a\otimes
  \mbox{D}Y^a\nn&&+\frac{\l^3}{4(1-\l^3 f^2)}\dd\r\otimes\dd\r\Big],
\eea
where
\bea
\l^3=\frac{2}{3(1+\r^2/8)},\;\; f^2=\frac{9}{16}\r^2;
\eea
 the $Y^a$, $a=1,...,4$ are constrained coordinates on an $S^3$,
 $Y^aY^a=1$, and we define
\bea
\mbox{D}Y^a=\dd Y^a+\omega^a_{\;\;\;b}Y^b,
\eea
where $\omega_{ab}$ are the spin connection one-forms of
$\mathcal{H}^4$. Finally $\dd s^2(\mathcal{H}^4)$ is normalised such
that the curvature two-form is given by $R_{ab}=-\frac{1}{3}e^a\w
e^b$. Defining the functions
\bea 
g=\sqrt{\frac{2}{3}}\frac{1}{\l^{1/2}m},\;\; h=\frac{\sqrt{1-\l^3
    f^2}}{\l^{1/2}m},
\eea
and with $\dd s^2(\mathcal{H}^4)=\d_{ab}e^a\otimes e^b$, the $SU(3)$
structure forms are given by 
\bea
e^7&=&-gY^ae^a,\nn
J&=&gh e^a\w \mbox{D}Y^a,\nn
\mbox{Re}\Omega&=&g^3\frac{1}{3!}\e^{abcd}Y^ae^b\w e^c\w
e^d-gh^2\frac{1}{2}\e^{abcd}Y^a\mbox{D}Y^b\w \mbox{D}Y^c\w e^d,\nn
\mbox{Im}\Omega&=&g^2h\frac{1}{2}\e^{abcd}Y^a\mbox{D}Y^b\w e^c\w
e^d-h^3\frac{1}{3!}\e^{abcd}Y^a\mbox{D}Y^b\w\mbox{D}Y^c\w\mbox{D}Y^d.
\eea

\subsection{Quaternionic K\"{a}hler geometry} 
In this case $\mathcal{M}_8$ admits a globally defined $Sp(2)$
structure. Defining the form $\Xi_1$ in terms of the three almost
complex structures according to
\bea
\Xi_1&=&\frac{1}{2}J^A\w J^A,
\eea
the $Sp(2)$ structure is realised by three globally defined null
Killing solutions of
\bea
\frac{1}{10}\G^{+-}\Xi_1\cdot\eta=-\eta.
\eea

\paragraph{Global truncation} The torsion conditions of the global
truncation are 
\bea
J^A\lrcorner\;\dd e^9&=&0,\nn
\dd (L^{-1}\mbox{Re}\Omega^A)&=&0,\nn
e^9\w[J^A\lrcorner \;\dd J^A-Le^9\lrcorner \;\dd(L^{-1}e^9)]&=&0,
\eea
where there is no sum on $A$ in the third
equation. Here $\Omega^A$ are the $(4,0)$ forms associated to the almost
complex structures $J^A$. More details of their definition are given in
the next section. The flux is
\bea
F&=&\dd(e^{+-9})+\frac{1}{3}\star \dd(e^{+-}\w
\Xi_1)+\frac{1}{6}L^{14/5}e^9\lrcorner\;\dd(L^{-14/5}\Xi_1)\nn&-&\frac{1}{4}\Xi_1\diamond[e^9\lrcorner\;(e^9\w\dd
e^9)]+F^{\mathbf{14}},
\eea
where $F^{\mathbf{14}}$ is a four-form on $\mathcal{M}_8$ in the
$\mathbf{14}$ of $Sp(2)$ which is unfixed by the truncation.

\paragraph{AdS geometry} The local metric for an $AdS_3$ 
region in this geometry is
\bea
\dd s^2=\frac{1}{\l m^2}\dd s^2(AdS_3) +e^A\otimes e^A+\hat{\rho}\otimes \hat{\rho}+\dd
s^2(\mathcal{N}_4),
\eea
where $\mathcal{N}_4$ admits a local $SU(2)$ structure, specified by a
triplet of self-dual $SU(2)$ forms $K^A$. The local $AdS_3$ torsion
conditions are
\bea
\hat{\r}\w\dd\left[\l^{-1}\left(\mbox{Vol}_4+\frac{1}{6}\e^{ABC}K^A\w
    e^{BC}\right)\right]&=&0,
\eea
\bea
\frac{1}{3}\left(K^A+\frac{1}{2}\e^{ABC}e^{BC}\right)\lrcorner\;\dd
e^A&=&\frac{2m\l^{1/2}}{\sin\theta}(2-\sin^2\theta)\nn&-&\cos\theta\hat{\r}\lrcorner\;\dd\log\left(\frac{\l^{3/2}\cos\theta}{\sin^2\theta}\right),\\
\dd[\l^{-3/2}\sin\theta(K^2\w e^2-K^1\w e^1)]&=&2m\l^{-1}[K^2\w
e^{31}-K^1\w e^{23}]\nn&+&2m\l^{-1}\cos\theta[K^2\w e^2-K^1\w e^1]\w\hat{\r},
\eea
together with permutations of the last equation. The magnetic flux is given by
\bea
F_{\mbox{\small{mag}}}&=&-\;\frac{\l^{3/2}}{\sin^2\theta}(\cos\theta+\star_8)\Big[\dd\Big(\l^{-3/2}\sin\theta\Big[\frac{1}{3}K^A\w
    e^A+e^{123}\Big]\Big)\nn&-&4m\l^{-1}\Big(\mbox{Vol}_4+\frac{1}{6}e^{ABC}K^A\w
e^{BC}\Big)\Big]-2m\l^{1/2}\Big[\frac{1}{3}K^A\w e^A+e^{123}\Big]\w\hat{\r}.\nn
\eea
Positive orientation on the space transverse to the $AdS$ factor is
defined by $\frac{1}{6}K^A\w K^A\w e^{123}\w\hat{\r}$.

\subsection{Complex Lagrangian geometry}
In this case, $\mathcal{M}_8$ again admits a global $Sp(2)$
structure. Defining
\bea
\Xi_2&=&\frac{1}{2}(J^1\w J^1-\mbox{Re}\Omega^2+\mbox{Re}\Omega^3),
\eea
it is realised by three globally defined null Killing solutions of
\bea
\G^{+-}\eta&=&\pm\eta,\\
\frac{1}{10}\G^{+-}\Xi_2\cdot\eta&=&-\eta.
\eea

\paragraph{Global truncation} In this case, effecting the global
truncation is technically more difficult. We have performed it under
the assumption that $e^9\w\dd e^9=0$. Then the torsion
conditions are given by
\bea
\dd(L^{-1/2}J^2)=\dd(L^{-1/2}J^3)&=&0,\nn
e^9\w[J^1\lrcorner \;\dd J^1-Le^9\lrcorner\;\dd(L^{-1}e^9)]&=&0.
\eea
The flux is 
\bea
F&=&\dd(e^{+-9})+\frac{1}{2}\star \dd(e^{+-}\w
\Xi_2)+\frac{1}{4}L^{11/5}e^9\lrcorner\;\dd(L^{-11/5}\Xi_2)+F^{\mathbf{14}}.
\eea

\paragraph{AdS geometry} The local frame, structure and orientation
for an $AdS_3$ region in this geometry are identical to
those in Quaternionic K\"{a}hler geometry. We have derived the $AdS_3$
torsion conditions by decomposing those in SLAG and K\"{a}hler-4
geometry (exactly how we do this is discussed in section 5) rather
than from the equations for the global truncation of the previous
paragraph. This means that our $AdS$ equations are independent of the
assumption $e^9\w\dd e^9=0$ that we made for the global Minkowski
frame above. The torsion conditions we find are
\bea
\hat{\r}\w\dd[\l^{-1}(\mbox{Vol}_4+K^3\w e^{12})]&=&0,\\
(K^3+e^{12})\lrcorner\;\dd
e^3&=&\frac{2m\l^{1/2}}{\sin\theta}(2-\sin^2\theta)-\cos\theta\hat{\r}\lrcorner\;\dd\log\left(\frac{\l^{3/2}\cos\theta}{\sin^2\theta}\right),\nn
\eea
\bea
\dd(\l^{-1}\sin\theta e^1)&=&m\l^{-1/2}(K^1+e^{23}+\cos\theta e^1\w\hat{\r}),\\
\dd(\l^{-1}\sin\theta e^2)&=&m\l^{-1/2}(K^2+e^{31}+\cos\theta e^2\w\hat{\r}).
\eea
The magnetic flux is
\bea
F_{\mbox{\small{mag}}}&=&\frac{\l^{3/2}}{\sin^2\theta}(\cos\theta+\star_8)[\dd(\l^{-3/2}\sin\theta[K^3\w
e^3+e^{123}])-4m\l^{-1}(\mbox{Vol}_4+K^3\w
e^{12})]\nn&+&2m\l^{1/2}[K^3\w e^3+e^{123}]\w\hat{\r}.
\eea

\section{Spinorial realisation of the frame bundles}
In this section, we will discuss the spinorial realisations of the
globally defined frame bundles we study. A global reduction of the frame bundle
to a sub-bundle is by definition equivalent 
to the existence of a globally-defined G-structure. The existence of a
globally defined G-structure, for our purposes, is equivalent to the
existence of a set of globally-defined forms, invariant under the action of the
structure group G. We will use the action of the structure forms on
the spin bundle to define the spinorial realisation of the
G-structure. When the flux vanishes asymptotically, the structure
forms asymptote to the calibrations of the asymptotic special holonomy
manifold.   

To start, we will specify the $Spin(1,10)$ structure we use for eleven
dimensional supergravity. We use all the supergravity and spinorial
conventions of \cite{j}, 
which are employed consistently throughout \cite{j}, \cite{j1}-\cite{oap1}, the
papers we will use in the next section for truncating supergravity to
the frame bundles of this section. We work in the null frame of the
introduction, 
\bea\label{fframe}
\dd s^2=2e^+\otimes e^-+\dd s^2(\mathcal{M}_8)+e^9\otimes e^9.
\eea
We recall that we impose, globally, that $e^+=L^{-1}\dd x^+$,
$e^-=\dd x^-$,  
$L<\infty$, $e^9\neq0$; and that $L$ and the frame on the space transverse to the
Minkowski factor are independent of the coordinates $x^{\pm}$. The
orientations we use are defined in section 2.

\subsection{$Spin(7)$ and associated local $AdS_3$  structures}
A global $Spin(7)$ structure in eleven dimensions is defined by the
no-where vanishing one-forms $e^{\pm}$, $e^9$, and the no-where 
vanishing Cayley four-form $\phi$. We choose the components of $\phi$ to be
\bea
-\phi&=&e^{1234}+e^{1256}+e^{1278}+e^{3456}+e^{3478}+e^{5678}+e^{1357}\nn&+&e^{2468}-e^{1368}-e^{1458}-e^{1467}-e^{2358}-e^{2367}-e^{2457}.
\eea
On a special holonomy manifold, $\phi$ calibrates Cayley
four-cycles. The embedding of our $Spin(7)$ structure group in
$Spin(1,10)$ (which is entirely at our discretion) is defined by
this choice of $\phi$, together with the globally-defined forms $e^{\pm}$,
$e^9$. 

The most general geometry we study is Cayley geometry, where the
$Spin(7)$ structure is realised by a single globally defined null
Killing spinor. This may be chosen to satisfy the projection
\bea
\frac{1}{14}\G^{+-}\phi\cdot\e&=&-\e.
\eea
With our choice of the components of $\phi$, this projection is
equivalent to
\bea\label{proj}
\G^{1234}\e=\G^{3456}\e=\G^{5678}\e=\G^{1357}\e=-\G^{+-}\e=-\e.
\eea
We will reserve the notation $\e$ for a globally-defined Killing
spinor satisfying this projection in the frame \eqref{fframe}. The
statements regarding the eigenvalues of the $Spin(7)$ modules of the
spin bundle may be verified by evaluating the Clifford action of
$\phi$ in a specific basis for the spin bundle; a useful choice (which
we will have used for all the spinor algebra described in this
section) is that constructed in \cite{ooo}. The spinor $\epsilon$ is
the spinorial
realisation of the frame bundle for the geometric dual of an $N=(1,0)$
CFT. 

Having found the spinorial realisation of the Cayley structure, the
structure forms may be obtained as bilinears of the
Killing spinor (apart from $e^-$, which is put in by hand in our
frame definition). As discussed in detail in \cite{j}, \cite{j1},
the only non-zero bilinears are the one-, two-, and five-forms, which
are
\bea
K&=&e^+,\nn
\Theta&=&e^{+9},\nn
\Sigma&=&e^+\wedge\phi.
\eea

An essential point in our construction is the patching of the
G-structures of the global $\mathbb{R}^{1,1}$ and local $AdS_3$ regions. We
will now examine this in detail, by imposing the most general local warped
product $AdS_3$ frame ansatz on our globally-defined
frame. Globally, we have
\bea\label{gg}
\dd s^2=L^{-1}\dd s^2(\mbb^{1,1})+\dd s^2(\mathcal{M}_8)+e^9\otimes
e^9.
\eea
Observe, that in Poincar\'{e} coordinates, every $AdS_3$ space is
foliated by $\mbb^{1,1}$ leaves:
\bea
\frac{1}{m^2}\dd s^2(AdS_3)=e^{-2mr}\dd s^2(\mbb^{1,1})+\dd r^2.
\eea
Therefore we demand that for a local $AdS$ region, $L$ in \eqref{gg}
is given by  
\bea
L=e^{2mr}\l,
\eea
for some function $\l$ which is independent of the $AdS$
coordinates. For a general $\mathbb{R}^{1,1}$ solution with an $AdS$
horizon, this expression for 
$L$ is local and valid for large positive $r$. For an $AdS_3$
conformal boundary, it is valid for large negative $r$. For a globally $AdS$
solution, it is valid for all $r$. To get an $AdS$ metric, we must
also pick out the $AdS$ radial one-form $\hat{r}=\l^{-1/2}\dd r$ from
the space transverse to the $\mbb^{1,1}$ factor. In an $AdS$ region,
this one-form will in general be a linear combination 
of $e^9$, and a one-form lying entirely in
$\mathcal{M}_8$. Using the transitive action of $Spin(7)$ on
$\mathcal{M}_8$ (an action which, by definition, leaves the Killing
spinor and Cayley form invariant) we may choose the part of $\hat{r}$
lying in $\mathcal{M}_8$ to lie entirely along the basis one-form
$e^8$. Then we may write the locally-defined $AdS_3$ frame as a
rotation of the globally-defined $\mbb^{1,1}$ frame,
as
\bea
 \hat{r}&=&\sin\theta e^8+\cos\theta e^9,\nn\label{a}
\hat{\rho}&=&\cos\theta e^8-\sin\theta e^9,
\eea
with\footnote{Obviously, $\theta=0$ is a special
  case; it will be discussed separately in section 5.} $0<\theta\le\pi/2$. We demand that $\hat{\rho}$, together with the remaining basis one-forms transverse
to the $AdS$ factor, are locally independent
of the $AdS$ coordinates. The local metric becomes
\bea
\dd s^2=\frac{1}{\l m^2}\dd s^2(AdS_3)+\hat{\rho}\otimes\hat{\rho}+ds^2(\mathcal{N}_7).
\eea
This frame-rotation technique was first employed in \cite{39}. Because we have locally picked out a preferred vector on
$\mathcal{M}_8$, the eleven-dimensional structure group is reduced,
locally, from $Spin(7)$ to a $G_2$ which acts
on $\mathcal{N}_7$. This $G_2$ structure is specified by the local
decomposition of the globally-defined $\phi$, into an associative
three-form $\Phi$ and a co-associative four-form $\Upsilon$ according to
\bea
-\phi=\Upsilon+\Phi\w e^8,
\eea
so that
\bea
\Phi&=&e^{127}+e^{347}+e^{567}+e^{246}-e^{136}-e^{145}-e^{235},\nn
\Upsilon&=&e^{1234}+e^{1256}+e^{3456}+e^{1357}-e^{1467}-e^{2367}-e^{2457}.
\eea

Having defined the most general global Minkowski and associated local
$AdS$ structures of interest to us, we now
describe how they may be further reduced, by imposing the existence of
even more exceptional global structures. The
first exceptional case we consider is where both spinorial singlets of
the global $Spin(7)$ structure are Killing. They are defined by  the projection
\bea
\frac{1}{14}\phi\cdot\eta=-\eta.
\eea
The local decomposition of the
Cayley four-form under $G_2$ in an $AdS_3$ patch will be exactly as
above; however, the local supersymmetry in the $AdS_3$ patch will
double, so now there will be four locally-defined $Spin(7)$
structures, whose common subgroup is the locally-defined $G_2$. The second
linearly independent globally defined Killing spinor realising the
maximal $Spin(7)$ structure is proportional to the basis
spinor \cite{ooo}
\bea
\G^{-}\e.
\eea
Now we will look at the spinorial realisations of frame bundles with a
reduced structure group.

\subsection{$SU(4)$ and associated local $AdS_3$  structures}
In this subsection we will define the spinorial realisations of an
$SU(4)$ frame bundle of interest to us. What we call $SU(4)$ geometry
is when all four 
spinorial singlets of the structure group of the frame bundle are
Killing and globally defined. The other spinorial realisations we
consider - defining what we call K\"{a}hler-4 and SLAG geometry - are
when two of the singlets of the structure group are Killing and
globally defined. Again, the Killing spinors may be
naturally selected by the action of the structure forms on the
spin bundle. With asymptotically vanishing flux, the interpolating
solutions of K\"{a}hler-4 and SLAG geometry will involve deformations of
the normal bundles of respectively K\"{a}hler-4 and SLAG-4 cycles of
the special holonomy manifolds to which they asymptote.

In the globally defined frame \eqref{fframe}, we 
demand that $\mathcal{M}_8$ admits an everywhere non-zero almost
complex structure 
two-form $J$ and a (4,0) form $\Omega$. We may always take
the $SU(4)$ structure group to be embedded in $Spin(1,10)$ such that
their components are given by  
\begin{eqnarray}
 J&=&e^{12}+e^{34}+e^{56}+e^{78}\;, \label{eqn:j4fold}\\
 \Omega&=&(e^1+ie^2)(e^3+ie^4)(e^5+ie^6)(e^7+ie^8)\;.
\end{eqnarray}

\paragraph{$SU(4)$ geometry} The $SU(4)$ singlets are defined globally by the action of the
structure forms on the spin bundle; they are the four solutions of
\bea
\G^{+-}\eta&=&\pm\eta,\\
\frac{1}{12}(J\wedge J)\cdot\eta&=&-\eta,
\eea
or equivalently
\bea
\G^{+-}\eta&=&\pm\eta,\\
\frac{1}{8}\mbox{Re}\Omega\cdot\eta&=&\pm\eta.
\eea
A third equivalent form of these conditions is
\bea
\G^{1234}\eta=\G^{3456}\eta=\G^{5678}\eta=\pm\G^{+-}\eta=-\eta\;.
\eea
$SU(4)$ geometry is defined by requiring that all four solutions of
these projections are Killing. Explicitly, the Killing solutions of
these equations are proportional to 
\bea\label{subspace}
\e,\;\G^-\e,\;\frac{1}{4}J\cdot\e,\;\frac{1}{4}\G^-J\cdot\e\;.
\eea
This maximal realisation of an $SU(4)$
structure is relevant for interpolations from Calabi-Yau cones to
Sasaki-Einstein manifolds. 

\paragraph{K\"{a}hler-4 geometry } 
The spinorial realisation of a K\"{a}hler-4 structure is given by two
globally defined Killing spinors which satisfy the projection 
\bea
\frac{1}{12}\G^{+-}(J\w J)\cdot\eta&=&-\eta.
\eea
This is equivalent to the maximal $SU(4)$ projections supplemented
by $\G^{+-}\eta=\eta$. The basis spinors of \eqref{subspace} which
survive this projection are
\bea
\e,\;\frac{1}{4}J\cdot\e\;. \label{eqn:skahler4}
\eea
Both of these spinors have positive chirality under $\G^{+-}$. The
bilinears associated to these basis spinors are
\bea
K&=&e^+,\nn
\Theta&=&e^{+9},\nn
\Sigma_{\pm}&=&e^+\w\phi_{\pm}\;, \label{eqn:kahler4}
\eea
where
\bea
\phi_{\pm}=-\left(\frac{1}{2}J\w J\pm\mbox{Re}\Omega\right)\;,
\eea
with $\phi_+$ coming from $\e$ and $\phi_-$ from $\frac{1}{4}J\cdot\e$.  
This is the spinorial realisation of the $SU(4)$ frame bundle of
relevance for the geometric duals of $N=(2,0)$ conformal field theories.

\paragraph{SLAG geometry} The spinorial realisation of the $SU(4)$
frame bundle defining SLAG geometry is specified by two
globally defined Killing spinors, which satisfy the projections
\bea
\G^{+-}\eta&=&\pm\eta,\nn
\frac{1}{8}\G^{+-}\mbox{Re}\Omega\cdot\eta&=&-\eta.
\eea
This is equivalent to the maximal $SU(4)$ projections,
supplemented by 
\bea
\G^{+-1357}\eta=-\eta.
\eea 
The pair of basis spinors surviving this projection are 
\bea
\e,\;\frac{1}{4}\G^-J\cdot\e \;. \label{eqn:sslag4}
\eea
The bilinears associated to the basis spinor
$\frac{1}{4}\G^-J\cdot\e$ are
\bea
K&=&-e^-,\nn
\Theta&=&e^{-9},\nn
\Sigma&=&e^-\w\left(\frac{1}{2}J\w J-\mbox{Re}\Omega\right). \label{eqn:slag4}
\eea
Observe that these Killing spinors have opposite $\mathbb{R}^{1,1}$
chirality, so this is 
the spinorial realisation of the $SU(4)$ frame bundle of relevance for
the geometric duals of $N=(1,1)$ CFTs. 

\paragraph{Local $AdS_3$ structures}
Now we will give the local $AdS_3$ structures which arise from each
spinorial realisation of the globally-defined $SU(4)$ structures. In this case, picking out
a local $AdS_3$ radial direction with a component on $\mathcal{M}_8$
reduces the structure group, locally, to $SU(3)$. The metric is given
by
\bea
\dd s^2=\frac{1}{\l m^2}\dd s^2(AdS_3)+e^7\otimes
e^7+\hat{\rho}\otimes\hat{\rho}+\dd s^2(\mathcal{N}_6),
\eea
where the locally defined $\mathcal{N}_6$ admits the local $SU(3)$
structure. For each of
$Spin(7)$ structures which collectively define a K\"{a}hler-4 structure, $-\phi_{\pm}=\Upsilon_{\pm}+\Phi_{\pm}\w e^8$, we find that the associated local
$AdS_3$ $G_2$
structures are 
\bea\label{kahler4}
\Phi_{\pm}&=&J_{SU(3)}\w e^7\mp\mbox{Im}\Omega_{SU(3)},\nn
\Upsilon_{\pm}&=&\frac{1}{2}J_{SU(3)}\w
J_{SU(3)}\pm\mbox{Re}\Omega_{SU(3)}\w e^7.
\eea
We see how the two local $G_2$ structures in turn define a local $SU(3)$
structure. For the globally-defined SLAG structures, the $G_2$
structures of a local $AdS_3$ patch are
\bea
\label{SLAG4}
\Phi_{\pm}&=&\pm J_{SU(3)}\w e^7-\mbox{Im}\Omega_{SU(3)},\nn
\Upsilon_{\pm}&=&\pm\frac{1}{2}J_{SU(3)}\w
J_{SU(3)}+\mbox{Re}\Omega_{SU(3)}\w e^7,
\eea
and collectively they define a different embedding of the local
$SU(3)$ in the local $G_2$. In
both cases 
\bea
J_{SU(3)}&=&e^{12}+e^{34}+e^{56},\nn
\Omega_{SU(3)}&=&(e^1+ie^2)(e^3+ie^4)(e^5+ie^6).
\eea
The local $AdS_3$ structure for $SU(4)$ geometry is obvious.

\subsection{$Sp(2)$ and associated local $AdS_3$ Structures}
Finally we will discuss the spinorial realistions of an $Sp(2)$ frame
bundle of interest to us. The discussion closely follows that of the $SU(4)$
case.

We obtain a globally defined $Sp(2)$ structure by demanding that
$\mathcal{M}_8$ admits a triplet of everywhere non-zero almost complex
structures $J^A$, $A=1,2,3$. These obey the algebra 
\bea
J^AJ^B=-\d^{AB}+\e^{ABC}J^C\;.
\eea
We can always choose a basis such that the components of the three almost complex are given by
\begin{subequations}
\begin{eqnarray}
J^1&=&e^{12}+e^{34}+e^{56}+e^{78}\;,\\
J^2&=&-e^{13}+e^{24}-e^{57}+e^{68}\;,\\
J^{3}&=&e^{14}+e^{23}+e^{67}+e^{58}\;.
\end{eqnarray}
\end{subequations}
Note that $J^1=J$, with $J$ given in \eqref{eqn:j4fold}. Each almost complex structure has a corresponding $(4,0)$ form given by
\bea
\Omega^1&=&\frac{1}{2}J^2\wedge J^2-\frac{1}{2}J^3\w J^3+iJ^2\w
J^3,\nn
\Omega^2&=&\frac{1}{2}J^3\wedge J^3-\frac{1}{2}J^1\w J^1+iJ^3\w
J^1,\nn
\Omega^3&=&\frac{1}{2}J^1\wedge J^1-\frac{1}{2}J^2\w J^2+iJ^1\w
J^2.
\eea

\paragraph{$Sp(2)$ geometry}
What we call $Sp(2)$ geometry is defined by the existence of six
Killing singlets of the structure group of the global $Sp(2)$ frame
bundle, which satisy the projections
\bea
\G^{+-}\eta&=&\pm\eta,\nn
\frac{1}{10}\Xi_1\cdot \eta&=&-\eta,
\eea
or equivalently 
\bea
\G^{+-}\eta&=&\pm\eta,\nn
\frac{1}{10}\Xi_2\cdot \eta&=&\pm\eta
\eea
The Killing solutions of these projections are proportional to the basis spinors
\bea
\e,\;\frac{1}{4}J^{1}\cdot\e,\;\frac{1}{4}J^{2}\cdot\e,\;
\G^-\e,\;\frac{1}{4}\G^-J^{1}\cdot\e,\;\frac{1}{4}\G^-J^{2}\cdot\e.
\eea
This realisation of an $Sp(2)$ frame bundle is of relevance for
interpolations from Hyperk\"{a}hler cones to Tri-Sasaki-Einstein manifolds.

\paragraph{QK geometry} QK geometry is defined by the existence of
three Killing spinorial realisations of the frame bundle which satisfy the projection
\bea
\frac{1}{10}\G^{+-}\Xi_1\cdot\eta=-\eta.
\eea
This projects out half of the Killing spinors of the maximal
structure; the Killing solutions of this projection are proportional
to the basis spinors
\bea
\e,\;\frac{1}{4}J^{1}\cdot\e,\;\frac{1}{4}J^{2}\cdot\e\;.
\eea
The bilinears associated to these basis spinors are
\bea
K^A&=&e^+,\nn
\Theta^A&=&e^{+9},\nn
\Sigma^A&=&e^+\wedge \phi^A, \label{eqn:qk4}
\eea
where 
\bea
\phi^A=-\frac{1}{2}J^A\wedge J^A-\mbox{Re}\Omega^A,
\eea
with no sum on $A$. This is the spinorial realisation of an $Sp(2)$
frame bundle of relevance to the geometric duals of $N=(3,0)$ CFTs. On
a special holonomy manifold, the supersymmetric cycle calibrated by $\Xi_1$ is
K\"{a}hler-4 with respect to all three complex structures.

\paragraph{CLAG geometry}  
CLAG geometry is defined by the existence of three Killing spinors
which satisfy the projections
\bea
\G^{+-}\eta&=&\pm\eta,\nn
\frac{1}{10}\G^{+-}\Xi_2\cdot\eta&=&-\eta.
\eea
Again, this projects out half
of the Killing spinors of the maximal structure; the subspace defined
by this projection is spanned by the basis spinors
\bea
\e,\;\frac{1}{4}J^1\cdot\e,\;\frac{1}{4}\G^-J^2\cdot\e.
\eea 

The bilinears associated to the first two of these basis spinors are
the same as in \eqref{eqn:qk4} with $A=1,3$; the bilinears associated
to $\frac{1}{4}\G^-J^2\cdot\e$ are
\bea
K&=&-e^-,\nn
\Theta&=&e^{-9},\nn
\Sigma&=&e^-\w\Big(\frac{1}{2}J^3\w J^3-\mbox{Re}\Omega^3\Big).
\eea
This is the spinorial realisation of the frame bundle of relevance to
the geometric duals of $N=(2,1)$ CFTs. On a special
holonomy manifold, the supersymmetric cycles calibrated by $\Xi_2$ are
K\"{a}hler-4 with repect to $J^1$, and SLAG with respect to
$-\mbox{Re}\Omega^2$ and $\mbox{Re}\Omega^3$.

\paragraph{Local $AdS_3$ structures}
In this case, picking out a local $AdS_3$ radial direction with a
component on $\mathcal{M}_8$ reduces the structure group near an $AdS$
horizon to $SU(2)$. The local metric is given by
\bea
\dd s^2=\frac{1}{\l m^2}\dd s^2(AdS_3) +e^5\otimes e^5+e^6\otimes
e^6+e^7\otimes e^7+\hat{\rho}\otimes \hat{\rho}+\dd
s^2(\mathcal{N}_4),
\eea
where the locally-defined $\mathcal{N}_4$ admits a local $SU(2)$
structure. For each of the three $Spin(7)$ structures which are
collectively equivalent to a
QK structure, $-\phi^A=\Upsilon^A+\Phi^A\w e^8$, we find the
associated local $AdS_3$ structures
\bea
\Phi^1&=&e^{567}+K^3\w e^7+K^2\w e^6-K^1\w e^5,\nn
\Upsilon^1&=&\mbox{Vol}_{\mathcal{N}_4}+K^3\w e^{56}-K^2\w e^{57}-K^1\w
e^{67},\nn\\
\Phi^2&=&e^{567}-K^3\w e^7+K^2\w e^6+K^1\w e^5,\nn
\Upsilon^2&=&\mbox{Vol}_{\mathcal{N}_4}-K^3\w e^{56}-K^2\w e^{57}+K^1\w
e^{67},\nn\\
\Phi^3&=&e^{567}+K^3\w e^7-K^2\w e^6+K^1\w e^5,\nn
\Upsilon^3&=&\mbox{Vol}_{\mathcal{N}_4}+K^3\w e^{56}+K^2\w e^{57}+K^1\w
e^{67},
\eea
where the $K^A$ are the self-dual $SU(2)$ invariant
two-forms on $\mathcal{N}_4$, given by
\bea
K^1&=&e^{14}+e^{23},\nn
K^2&=&-e^{13}+e^{24},\nn
K^3&=&e^{12}+e^{34}.
\label{eqn:ks}
\eea
They satisfy the algebra $K^AK^B=-\delta^{AB}-\e^{ABC}K^C$. For CLAG
geometry, two of 
the local $G_2$ structures $\{\Phi^1,\Upsilon^1\}$,
$\{\Phi^3,\Upsilon^3\}$, have exactly the same form as for 
QK geometry, while the third structure is now given by
\bea
\Phi^2&=&-e^{567}+K^1\w e^7-K^2\w e^6-K^3\w e^5,\nn
\Upsilon^2&=&-\mbox{Vol}_{\mathcal{N}_4}+K^1\w e^{56}+K^2\w e^{57}-K^3\w e^{67}.
\eea

\section{Truncating eleven-dimensional supergravity}
In this section, we will truncate eleven dimensional supergravity to
the global frame bundles of interest to us. The different ways in
which we do this are parameterised by the different spinorial
realisations of the frame bundles we defined in the previous section. The papers
\cite{j}, \cite{j1}-\cite{oap1} essentially provide a machine for
doing this. The input is the global Minkowski frame, and in each
case, the most general Killing spinors satisfying the appropriate global
projection conditions. The output (with human intervention) is the most
general BPS conditions in each case. These, coupled to the Bianchi
identity and outstanding component of the field equations, define the
truncation of eleven dimensional supergravity to the frame bundles.

Since they are qualitatively similar to one another (and qualitatively
different to the other cases), and have also already received much
attention, we will first briefly discuss the maximal structures,
before moving on to the remaining cases.

\subsection{$Spin(7)$, $SU(4)$ and $Sp(2)$ geometry}
The BPS conditions for the maximal structures may be obtained by a
trivial restriction and globalisation of the local conditions of
\cite{oo} (for $Spin(7)$) and \cite{oap} (for 
$SU(4)$ and $Sp(2)$) to our global $\mathbb{R}^{1,1}$ frame. To state
them, it is convenient to make some frame redefinitions (for this
subsection only) so let us define $L=H^{2/3}$, $e^9=H^{-1/3}\hat{e}^9$, and
conformally rescale the frame on $\mathcal{M}_8$ so that the metric
becomes
\bea
\dd s^2=H^{-2/3}[\dd
s^2(\mbb^{1,1})+\hat{e}^9\otimes\hat{e}^9]+H^{1/3}\dd
\tilde{s}^2(\mathcal{M}_8).
\eea
The Killing spinors for $Spin(7)$ geometry are given by
\bea
\e,\;\;H^{-1/3}\G^-\e.
\eea
The torsion conditions in this case may be succinctly summarised by saying that
$\hat{e}^9$ is Killing, $\dd \hat{e}^9$ is a two-form on
$\mathcal{M}_8$ in the $\rep{21}$ of $Spin(7)$, and
$\dd\tilde{s}^2(\mathcal{M}_8)$ is globally a metric of $Spin(7)$
holonomy. The flux is given by
\bea
F=\dd (e^{+-9})+F^{\mathbf{27}},
\eea
where $F^{\mathbf{27}}$ is a four-form on $\mathcal{M}_8$ in the
$\rep{27}$ of $Spin(7)$. The Bianchi identity and field equation
reduce to
\bea
\dd F^{\rep{27}}&=&0,\\
\tilde{\nabla}^2H&=&-\dd\hat{e}^9\lrcorner\dd\hat{e}^9-\frac{1}{2}F^{\rep{27}}\lrcorner
F^{\rep{27}},
\eea
where in the second equation all operations are defined in the conformally
rescaled metric. For $SU(4)$ geometry,
$\dd\tilde{s}^2(\mathcal{M}_8)$ is globally restricted further to a
metric of $SU(4)$ holonomy, $\dd \hat{e}^9$ to the $\rep{15}$, and
$F_{\mbox{\small{mag}}}$ to the $\mathbf{20}$. For $Sp(2)$ geometry,
$\dd\tilde{s}^2(\mathcal{M}_8)$ has global $Sp(2)$ holonomy, $\dd \hat{e}^9$
belongs to the $\rep{10}$, and $F_{\mbox{\small{mag}}}$ to the $\mathbf{14}$. The
most general $AdS_3$ horizons in these truncations are the
Freund-Rubin solutions, the direct products
$AdS_4\times\mathcal{M}_7$. In $Spin(7)$ geometry,
$\mathcal{M}_7$ has weak $G_2$ holonomy; for $SU(4)$ geometry, $\mathcal{M}_7$ is Sasaki-Einstein; and for 
$Sp(2)$ geometry, it is Tri-Sasaki-Einstein. These geometries were
discussed in detail in \cite{jose}. We have nothing else to
say about maximal structures, and will henceforth focus on the
remaining cases, where the global frame bundles are realised by half
the number of Killing spinors as the maximal structures.

\subsection{Cayley geometry}
Cayley geometry is given by a globally defined $Spin(7)$ frame
bundle realised by a single globally defined global null Killing
spinor. In \cite{j}, the most general local BPS conditions implied by
the existence of a single locally defined null Killing spinor were
derived. We may thus obtain the truncation of eleven dimensional
supergravity to Cayley geometry simply by restricting the
conditions of \cite{j} to a Minkowski frame and globalising them. The single null Killing spinor is $\e$, and the torsion
conditions and flux are as given in the introduction.

\subsection{K\"{a}hler-4, SLAG, QK and CLAG  geometry}
Now we move to the remaining cases, where the
derivation of the torsion conditions is considerably more involved. We
have used a combination of techniques. The K\"{a}hler-4 and QK
torsion conditions may be extracted, with considerable effort, by
restricting and globalising
the appropriate local classifications of \cite{oap}. We derive the SLAG
and CLAG conditions from scratch, using the machinery of \cite{oap1}. In the
K\"{a}hler-4 case, the general solution for the Killing spinors, given
our frame, is
\bea
\e;\;\;\frac{1}{4}J\cdot\e.
\eea
The Killing spinors have constant components in this spinorial
basis. The same is true of the QK Killing spinors; in general they
are
\bea
\e;\;\;\frac{1}{4}J^1\cdot\e;\;\;\frac{1}{4}J^2\cdot\e.
\eea
For SLAG geometry, the general solution for the Killing spinors is
\bea
\e;\;\;L^{-1/2}\frac{1}{4}\G^-J\cdot\e.
\eea
Finally for CLAG, the Killing spinors are
\bea
\e;\;\;\frac{1}{4}J^1\cdot\e;\;\;L^{-1/2}\frac{1}{4}\G^-J^2\cdot\e.
\eea
In all cases, a useful consistency check on our torsion conditions
and expressions for the flux is provided by the generalised
calibration conditions of \cite{j1}: 
\bea
\dd K&=&\frac{2}{3}\Theta\lrcorner F+\frac{1}{3}\Sigma\lrcorner\star F,\nn
\dd \Theta&=&K\lrcorner F,\nn
\dd\Sigma&=&K\lrcorner\star F-\Theta\wedge F.\label{cal}
\eea 
These are conditions on the exterior derivatives of the bilinears of
the Killing spinors, in eleven dimensions. For any of the Killing
spinors we look at, with a flux of the form of \eqref{1.5}, these are
equivalent to
\bea
H&=&L\dd(L^{-1}e^9),\nn
\dd \log
L&=&\frac{2}{3}Le^9\lrcorner \;\dd(L^{-1}e^9)+\frac{1}{3}\phi\lrcorner\star_9G,\nn
L\dd(L^{-1}\phi)&=&-\star_9G+e^9\w G.
\eea
The particular choice of $\phi$ depends on the particular choice of
Killing spinor, and is as given in the previous section. A module
of the flux which is in fact fixed by supersymmetry drops out of
these equations, so they are not sufficient conditions for
supersymmetry in general. However, for the modules which they contain,
they provide us with a useful consistency check. The results we obtain
are as stated in section two. The details of the calculations are
uninstructive, how to do them is explained in \cite{oap}, \cite{oap1},
and so we have suppressed them.

\section{The $AdS_3$  geometries}
In this section, we will explain in more detail how we derive the
conditions on the 
geometry of $AdS$ boundary regions of solutions of
Cayley, K\"{a}hler-4, SLAG, QK and CLAG geometry. By boundary, we mean
either event horizon or conformal boundary. For
Cayley geometry, we derive the conditions by inserting the local $AdS_3$ frame of 3.1 into
the torsion conditions for the global truncation, and also by demanding
that the flux at an $AdS$ boundary respects the $AdS$
isometries. This requirement on the flux means that it takes the form
\bea
F=\mbox{Vol}_{AdS_3}\w g+F_{\mbox{\small{mag}}},
\eea
with
\bea
\partial_{\hat{a}}g=\partial_{\hat{a}}F_{\mbox{\small{mag}}}=\hat{e}^a\lrcorner\;F_{\mbox{\small{mag}}}=0,
\eea
where $\hat{e}^a$ are the $AdS$ basis one-forms. As explained in
section three, the global structure group 
is reduced locally at an $AdS$ boundary. We define the $AdS$
geometry by the conditions on the intrinsic torsion of
the locally defined structure, together with the flux, in each
case.

An interesting question is that of the global structure of a manifold
with global $AdS_3$ isometry. Recall the frame rotation of 3.1:   
\bea\label{aaaa}
 \hat{r}&=&\sin\theta e^8+\cos\theta e^9,\nn
\hat{\rho}&=&\cos\theta e^8-\sin\theta e^9,
\eea
with $\hat{r}=\l^{-1/2}\dd r$ and $L=e^{2mr}\l$. In the generic case
of a Cayley bundle, provided that $\theta\neq0$ 
globally, this frame rotation reduces the global 
$Spin(7)$ structure associated to the $\mathbb{R}^{1,1}$
isometry to a global $G_2$ structure associated to the $AdS_3$
isometry. Thus for a manifold with global $AdS_3$ isometries, the
$G_2$ structure will only fail to be globally defined if there exist
points where $\theta=0$. It may be readily verified that the torsion
conditions and flux for a Cayley bundle imply that $\theta=0$ in an
open neighbourhood is
inconsistent with $AdS_3$ isometry of the neighbourhood - an $AdS_3$ frame and flux with
$\hat{r}=e^9$ does not solve the supergravity equations for a Cayley frame
bundle. A much more subtle issue - which we have not attempted to
resolve - is what happens at isolated points of a global $AdS$
manifold where $\theta=0$, and what the existence of such points
implies for the geometry or topology.

In the remainder of this section, we will first discuss how, by imposing $AdS_3$
isometry on the electric flux (which is universally given by
$F_{\mbox{\small{elec}}}=\dd(e^{+-9})$ in all cases we study) one may
introduce local coordinates for the local $AdS$ frame. We will then
show how the general necessary and sufficient 
minimally supersymmetric $AdS_3$ conditions of Martelli and Sparks may
be derived by imposing an $AdS_3$ boundary condition on an
arbitrary solution of Cayley geometry. Finally we will derive
the $AdS$ boundary geometry of a solution with a
K\"{a}hler-4, SLAG, QK or CLAG bundle.

\subsection{Coordinates for the $AdS$ frame}
\label{sec:coordsAdS}
Imposing $AdS_3$ isometries on the electric flux, we demand that
\bea
L\dd(L^{-1}e^9)=\hat{r}\wedge g,
\eea
with $g$ independent of $r$. Then using (\ref{aaaa}), we get
\bea
e^{2mr}\partial_r(e^{-2mr}\sin\theta)\l^{1/2}\hat{\r}+\l^{3/2}\tilde{\dd}(\l^{-3/2}\cos\theta)&=&-g,\nn
\tilde{\dd}(\l^{-1}\sin\theta\hat{\r})&=&0,
\eea
where $\tilde{\dd}$ denotes the exterior derivative on the space
transverse to the $AdS$ factor. Since $g$, $\hat{\r}$ and $\l$ are independent
of $r$, the first of these equations implies
that the rotation angle $\theta$ is also independent of $r$. Then the
second equation implies that locally there exists a coordinate $\rho$ such that
\bea
\hat{\r}=\frac{\l}{2m\sin\theta}\dd\r.
\eea
Therefore in general, we may write the metric near an $AdS_3$ boundary
as
\bea
\dd s^2=\frac{1}{\l m^2}\left[\dd
  s^2(AdS_3)+\frac{\l^3}{4\sin^2\theta}\dd\r\otimes\dd\r\right]+\dd
s^2(\mathcal{N}_7),
\label{eqn:genAdS}
\eea
where $\mathcal{N}_7$ is defined by
\bea
\dd s^2(\mathcal{M}_8)=\dd s^2(\mathcal{N}_7)+e^8\otimes e^8.
\eea

There is a special case in which we can do more, and integrate
the frame rotation completely. In general, in the Minkowski frame, there
exists a coordinate $z$  such that the one-form $e^9$ is given by
\bea
e^9=C(\dd z+\s),
\eea
for a function $C$ and a one-form $\s$ on $\mathcal{M}_8$ which are
independent of the Minkowski coordinates. When $\s=0$ (equivalently,
when $e^9\w\dd e^9=0$) both the
Minkowski and associated $AdS$ BPS conditions simplify considerably,
and we can integrate the frame rotation. 

To see how to do this, we first solve for $\dd z$ in \eqref{aaaa},
and then take the exterior derivative. We get
\bea
\dd(\l^{-1/2}C^{-1}\cos\theta)\w \dd r-\frac{1}{2m}\dd(\l C^{-1})\wedge \dd\r=0.
\eea
We may immediately deduce that $C=\tilde{C}(r,\r)\l$. Then
\bea
\dd(\l^{-3/2}\tilde{C}^{-1}\cos\theta)\w \dd
r=\frac{1}{2m}\partial_r\tilde{C}^{-1}\dd r\w \dd\r
\eea
implies that 
\bea
\l^{-3/2}\cos\theta=f(\r),
\eea
for some arbitrary function $f$. Therefore, when $\s=0$ in the
Minkowski frame, we may write the metric in the $AdS$ frame as
\bea\label{b}
\dd s^2=\frac{1}{\l
  m^2}\Big[\dd s^2(AdS_3)+\frac{\l^3}{4(1-\l^3f^2)}\dd\r\otimes\dd\r\Big]+ds^2(\mathcal{N}_7).
\eea
The electric flux is then given by
\bea
g=(1-\partial_{\r}f)\l^{3/2}\dd\r.
\eea
It is instructive to compare this expression with the results of
\cite{wrap}, where the supersymmetry conditions for $AdS$ boundaries
with global Minkowski frames, frame bundles with structure group
contained in $G_2$, and purely magnetic fluxes, were
derived. For the $AdS$ geometries, an expression for the metric of the form of
(\ref{b}) was found, in every case with $f=\r$. In the present
context, we see that when $\s=0$, $f$ essentially sources the electric flux, and
that as in \cite{wrap}, $f=\r$ implies that the fluxes are purely
magnetic. To conclude this subsection, we record the expression for $e^8$ in the
$AdS$ frame, when $\s=0$ in the Minkowski frame:
\bea\label{eqn:e8}
e^8=\l^{-1/2}\sqrt{1-\l^3f^2}\dd r+\frac{\l^{5/2}f}{2m\sqrt{1-\l^3f^2}}\dd\r.
\eea

\subsection{$AdS$ boundaries in Cayley geometry}
In this subsection, we will derive the minimal $AdS_3$ BPS conditions
quoted in the introduction by imposing an $AdS_3$ boundary
condition on Cayley geometry. First we will derive the $AdS$
torsion conditions, and then the relationship between the flux and the
torsion in the $AdS$ limit. We start with the torsion conditions 
\bea\label{A1}
e^9\w\left[-L^3e^9\lrcorner \dd
  (L^{-3}e^9)+\frac{1}{2}\phi\lrcorner\dd\phi\right] &=&0,\\\label{A2}
(e^9\w+\star_9)[e^9\lrcorner\dd (L^{-1}\phi)]&=&0.
\eea
One of the reasons why we have written the torsion conditions in this coordinate
independent form is that it makes it much easier to perform the frame
rotation. Now we do this, using
\bea
\phi&=&-\Upsilon-\Phi\w e^8,\nn
e^8&=&\sin\theta \hat{r}+\cos \theta\hat{\r},\nn
e^9&=&\cos\theta\hat{r}-\sin\theta\hat{\r},
\eea
to evaluate \eqref{A1} and \eqref{A2} in the $\hat{r}$, $\hat{\r}$
frame. We have seen that $\theta$ must be independent of the $AdS$ radial
coordinate, and we demand that the only $r$ dependence in the rotated frame
enters in the warping of the $\mathbb{R}^{1,1}$ factor. We split all exterior derivative as $\dd
=\hat{r}\w\partial_{\hat{r}}+\hat{\r}\w\pr+\dn$. Separating out the
$\hat{r}\w \hat{\r}$ and $\cos\theta\hat{r}-\sin\theta\hat{\r}$ components,
\eqref{A1} contains the two independent equations
\bea\label{A5}
\frac{1}{2}\Upsilon\lrcorner\dn\Phi+\frac{1}{2}\cos\theta\Upsilon\lrcorner\pr\Upsilon+\l^{7/2}\pr(\l^{-7/2}\cos\theta)
-6m\l^{1/2}\sin\theta&=&0,\\\label{A6}
\Upsilon\lrcorner\dn\Upsilon-\frac{1}{2}\cos\theta\Phi\lrcorner\pr\Upsilon-3\dn\log\l+\cos^2\theta\dn\log\left(\frac{\l^{3/2}\cos\theta}{\sin^2\theta}\right)&=&0,
\eea
where in deriving the second of these equations we have used the $G_2$
identities $\Phi\lrcorner\dn\Phi=-\Upsilon\lrcorner\dn\Upsilon$,
$\Phi\lrcorner(A\w\Phi)=-4A$. Applying the same procedure to
\eqref{A2}, we find the single condition
\bea
0&=&4m\l^{1/2}\cos\theta\Upsilon+\l^{3/2}\star_7\pr(\l^{-3/2}\sin\theta\Phi)+\l\sin\theta\pr(\l^{-1}{\Upsilon})\nn&+&\label{A7}\sin\theta\cos\theta\dn\log\left(\frac{\l^{3/2}\cos\theta}{\sin^2\theta}\right)\w\Phi.
\eea

To proceed, we decompose \eqref{A5}, \eqref{A6} and \eqref{A7} into
modules of $G_2$, to extract out the independent conditions. We will
then show that these can be repackaged in the form of
\cite{J+D}. First consider \eqref{A7}. This is an equation for
four-forms of $G_2$, and hence a priori contains $\rep{1}$, $\rep{7}$ and $\rep{27}$
parts. To treat the $\pr$ terms, it is useful to introduce
\bea
Q_{ij}=\d_{ik}(\pr e^k)_j,
\eea
where indices run from 1 to 7. Since we have chosen the frame so that $\Phi$ and $\Upsilon$ have
constant components, we may then write
\bea
(\pr \Phi)_{i_1i_2i_3}&=&3\Phi_{k[i_1i_2}Q^k_{\;\;\;i_3]},\nn\label{A8}
(\pr
\Upsilon)_{i_1i_2i_3i_4}&=&-4\Upsilon_{k[i_1i_2i_3}Q^k_{\;\;\;i_4]}.
\eea
Since $Q$ is an a priori arbitrary rank 2 tensor of $G_2$,
it contains $\rep{1}$, $\rep{7}$, $\rep{14}$ and $\rep{27}$ parts, and
encodes the intrinsic torsion modules of the eight-dimensional $G_2$
structure contained in the $\hat{\r}$ derivatives of $\Phi$ and
$\Upsilon$. Acting on these $G_2$ invariant forms, the $\rep{14}$ part
of $Q$ drops out of \eqref{A8}. We can separate out the remaining
parts of $Q$ according to
\bea
Q_{ij}=\frac{1}{7}\gamma\d_{ij}+\Phi_{ijk}\beta^k+Q_{ij}^{\rep{27}},
\eea
where $Q_{ij}^{\mathbf{27}}$ is a symmetric traceless tensor. Now
we insert this expression for $Q_{ij}$, together with \eqref{A8},
into \eqref{A7}. The $Q_{ij}^{\rep{27}}$ drops out; this is most
easily seen by choosing any particular element of the $\rep{27}$ and
verifying that its contribution to \eqref{A7} vanishes. The remaining
terms are given by 
\bea
0&=&\left[4m\l^{1/2}\cos\theta+\l^{5/2}\pr(\l^{-5/2}\sin\theta)+\sin\theta\gamma\right]\Upsilon\nn&+&\left[\sin\theta\cos\theta\dn\log\left(\frac{\l^{3/2}\cos\theta}{\sin^2\theta}\right)-6\sin\theta\beta\right]\w\Phi,
\eea
where in evaluating the $\rep{7}$ terms we have used the $G_2$
identities given in the appendix of \cite{bilal}. Therefore we must
have
\bea\label{A9}
\gamma&=&-4m\l^{1/2}\frac{\cos\theta}{\sin\theta}-\pr\log(\l^{-5/2}\sin\theta),\\\label{A10}
\beta&=&\frac{1}{6}\cos\theta\dn\log\left(\frac{\l^{3/2}\cos\theta}{\sin^2\theta}\right).
\eea 
This exhausts the content of \eqref{A7}. Now, using our expression for
$\gamma$, \eqref{A5} gives the singlet part of $\dn\Phi$. We find
\bea\label{A11}
\Upsilon\lrcorner\dn\Phi&=&\frac{4m\l^{1/2}}{\sin\theta}(4-\sin^2\theta)-2\cos\theta\pr\log\left(\frac{\l^{3/2}\cos\theta}{\sin^2\theta}\right).
\eea
Next, using our expression for $\beta$ in \eqref{A6}, we obtain 
\bea\label{A12}
\Upsilon\lrcorner\dn\Upsilon=3\dn\log\l.
\eea
This exhausts all the torsion conditions. Finally, it may be verified
that the four conditions \eqref{A9}-\eqref{A12} are equivalent to
\bea\label{A13}
\hat{\r}\w\dd(\l^{-1}\Upsilon)^{\rep{7}}&=&0,\\\label{A14}
\l^{5/2}\dd\left(\l^{-5/2}\sin\theta\mbox{Vol}_7\right)&=&-4m\l^{1/2}\cos\theta\hat{\r}\w\mbox{Vol}_7,\\
\dd\Phi\w\Phi&=&\frac{4m\l^{1/2}}{\sin\theta}(4-\sin^2\theta)\mbox{Vol}_7-2\cos\theta\star_8\dd\log\left(\frac{\l^{3/2}\cos\theta}{\sin^2\theta}\right).\nn\label{A15}
\eea
Equation \eqref{A13} is equivalent to \eqref{A12}, \eqref{A14} to
\eqref{A9}, and \eqref{A15} is equivalent to \eqref{A10} and \eqref{A11}. 

Next we must impose the $AdS$ boundary condition on the magnetic
flux, which in the Minkowski frame is
\bea
F_{\mbox{\small{mag}}}=-\star\dd(e^{+-}\w\phi)-\frac{L^{10/7}}{2}e^9\lrcorner\;\dd(L^{-10/7}\phi)+\frac{1}{4}\phi\diamond[e^9\lrcorner\;(e^9\w\dd
e^9)]+F^{\rep{27}}.
\eea
We define 
\bea
F^{\rep{27}}=G^{\rep{27}}+H^{\rep{27}}\w e^8;
\eea
eight-dimensional self-duality of $F^{\rep{27}}$ then implies that
$G^{\rep{27}}=\star_7H^{\mathbf{27}}$ (recall that the $\rep{27}$ of
$Spin(7)$ is irreducible under $G_2$). Now we perform the frame
rotation, and impose vanishing of the components of $F_{\mbox{\small{mag}}}$
along the $AdS$ radial direction. We find the conditions
\bea\label{A19}
\hat{\rho}\w\dd(\l^{-1}\Upsilon)&=&0,\\
H^{\rep{27}}&=&\frac{1}{\sin\theta}\star_7\l\pr
(\l^{-1}\Upsilon)+\star_7\dn\left(\frac{\cos\theta}{\sin\theta}\Phi\right)+\frac{10m}{7}\l^{1/2}\cos\theta\Phi\nn&
+&\frac{1}{2}\l^{27/14}\pr(\l^{-27/14}\sin\theta\Phi)+\frac{1}{4}\sin\theta\cos\theta\dn\log(\l^{-3/2}\cos\theta)\lrcorner\Upsilon.\nn\label{A20}
\eea
The first of these equations implies \eqref{A13}, and together with
\eqref{A14} and \eqref{A15}, comprises the torsion conditions given in
the introduction. The left-hand side of \eqref{A20} only contains a
term in the 
$\rep{27}$ of $G_2$, and hence the $\mathbf{1}$ and $\rep{7}$ parts of the
right-hand side must vanish. It may be verified that they do, using
\eqref{A13}, \eqref{A14} and \eqref{A15}. Using \eqref{A20}, the
magnetic flux may be expressed as
\bea
F_{\mbox{\small{mag}}}&=&\hat{\r}\w\Big[-\frac{\cos\theta}{\sin\theta}\star_7\l\pr(\l^{-1}\Upsilon)-\cos\theta\star_7\dn\left(\frac{\cos\theta}{\sin\theta}\Phi\right)\nn&&+2m\l^{1/2}\Phi-\star_7\l^{3/2}\dn(\l^{-3/2}\sin\theta\Phi)\Big]\nn
&&+\Big[\star_7H^{\rep{27}}-\frac{24m}{7}\l^{1/2}\cos\theta\Upsilon-\star_7\l^{3/2}\pr(\l^{-3/2}\sin\theta\Phi)\nn
&&-\frac{1}{2}\sin\theta\l^{10/7}\pr(\l^{-10/7}\Upsilon)+\frac{1}{4}\sin\theta\cos\theta\dn\log\left(\frac{\sin^4\theta}{\l^{9/2}\cos\theta}\right)\w\Phi\Big].
\eea
After some manipulation, this expression may be shown to be equivalent
to
\bea\label{nme}
F_{\mbox{\small{mag}}}=\frac{\l^{3/2}}{\sin^2\theta}\Big(\cos\theta+\star_8\Big)\Big(\dd[\l^{-3/2}\sin\theta\Phi]-4m\l^{-1}\Upsilon\Big)+2m\l^{1/2}\Phi\w\hat{\r},
\eea
which is in turn equivalent to
\bea
\l^{3/2}\dd(\l^{-3/2}\sin\theta\Phi)=\star_8F_{\mbox{\small{mag}}}-\cos\theta
F_{\mbox{\small{mag}}}+2m\l^{1/2}(\Upsilon+\cos\theta\Phi\w\hat{\r}).
\eea
This exhausts all conditions.

\subsection{$AdS$ boundaries in K\"{a}hler-4 and SLAG geometry}
There are two ways in which we can derive the BPS conditions for an $AdS$
boundary in K\"{a}hler-4 or SLAG geometry. The first is to
impose the frame rotation on the appropriate global truncation of
eleven dimensional supergravity, just as for Cayley geometry. The second and technically simpler way is to use the
local $AdS$ structures of section 3. This is what we have done. An
$AdS$ region in K\"{a}hler-4 geometry admits a pair of local
$G_2$ structures, which are equivalent to a local $SU(3)$ structure,
according to
\bea
\Phi_{\pm}&=&J_{SU(3)}\w e^7\mp\mbox{Im}\Omega_{SU(3)},\nn
\Upsilon_{\pm}&=&\frac{1}{2}J_{SU(3)}\w
J_{SU(3)}\pm\mbox{Re}\Omega_{SU(3)}\w e^7.
\eea
Both these $G_2$ structures must satisfy the local Cayley $AdS_3$
conditions. Similarly for SLAG geometry, where the local $G_2$
structures of an $AdS$ region are  
\bea
\Phi_{\pm}&=&\pm J_{SU(3)}\w e^7-\mbox{Im}\Omega_{SU(3)},\nn
\Upsilon_{\pm}&=&\pm\frac{1}{2}J_{SU(3)}\w
J_{SU(3)}+\mbox{Re}\Omega_{SU(3)}\w e^7,
\eea
In each case, the local $AdS$ metric is 
\bea
\dd s^2=\frac{1}{\l m^2}\Big[\dd s^2(AdS_3)+\frac{\l^3}{4\sin^2\theta}\dd\r\otimes\dd\r\Big]+e^7\otimes
e^7+\dd s^2(\mathcal{N}_6).
\eea
For the remainder of this subsection it is understood that structure
forms are of $SU(3)$, and we will suppress their subscripts. This
doubling of the structures provides a very convenient way of arriving
at the BPS conditions in each case. For example, the Cayley condition
$\hat{\r}\w\dd(\l^{-1}\Upsilon)=0$ decomposes for both K\"{a}hler-4 and
SLAG into the pair of equations
\bea
\hat{\rho}\w\dd(\l^{-1}J\w J)&=&0,\nn  
\hat{\rho}\w\dd(\l^{-1}\mbox{Re}\Omega\w e^7)&=&0.
\eea
Similarly, decomposing \eqref{A14} and \eqref{A15} leads to identical
equations for $AdS_3$ regions in both K\"{a}hler-4 and SLAG
geometries. What distinguishes these geometries is the decomposition
of the flux. Requiring a magnetic flux of the form \eqref{nme} for both the
local $G_2$ structures in K\"{a}hler-4 geometry, we find the
conditions
\bea
F_{\mbox{\small{mag}}} &=&  \frac{\l^{3/2}}{\sin^2\theta}(\cos\theta+\star_8)(\dd[\l^{-3/2}\sin\theta
J\w e^{7}
]-2m\l^{-1} J\w J  ) + 2m\l^{1/2} J\w  e^{7}\w \hat{\r},\nn\\
0 &=&  -\;\frac{\l^{3/2}}{\sin^2\theta}(\cos\theta+\star_8)(\dd[\l^{-3/2}\sin\theta
\mbox{Im}\Omega
]+4m\l^{-1} \mbox{Re}\Omega\w e^7 )  - 2m\l^{1/2} \mbox{Im}\Omega\w \hat{\r}.\nn
\eea
For the local $AdS$ structures in SLAG geometry, we instead get
\bea
0&=&  \frac{\l^{3/2}}{\sin^2\theta}(\cos\theta+\star_8)(\dd[\l^{-3/2}\sin\theta
J\w e^{7}
]-2m\l^{-1} J\w J  ) + 2m\l^{1/2} J\w  e^{7}\w \hat{\r},\nn\\
F_{\mbox{\small{mag}}}&=&-\;\frac{\l^{3/2}}{\sin^2\theta}(\cos\theta+\star_8)(\dd[\l^{-3/2}\sin\theta\mbox{Im}\Omega]+4m\l^{-1}\mbox{Re}\Omega\w
e^7)-2m\l^{1/2}\mbox{Im}\Omega\w\hat{\r}.\nn
\eea
It is this formal difference in the decomposition of the flux that
endows $AdS$ regions in the two geometries with such different
properties. 

In each case, not all the equations obtained by performing this
decomposition of the Cayley $AdS$ conditions are independent. We have
reduced them to a minimal set of necessary and sufficient independent
conditions, which are quoted in section 2.

\subsection{$AdS$ boundaries in QK and CLAG geometry}
To derive the BPS conditions for an $AdS$ boundary in QK or CLAG
geometry, we may either impose the frame rotation on the torsion
conditions of section 4, or (which is again technically more
convenient) we can use the local $AdS_3$ structures of section 3 to further decompose
the torsion conditions for an $AdS$ region in K\"{a}hler-4 or SLAG
geometry. The derivation proceeds in a very similar way to that of
the derivation of the geometry of $AdS$ regions in K\"{a}hler-4 or
SLAG from Cayley, and we have suppressed the details. In QK and CLAG
geometry, the local $AdS$ metric is  
\bea
\dd s^2=\frac{1}{\l m^2}\Big[\dd s^2(AdS_3)+\frac{\l^3}{4\sin^2\theta}\dd\r\otimes\dd\r\Big] +e^5\otimes e^5+e^6\otimes
e^6+e^7\otimes e^7+\dd
s^2(\mathcal{N}_4).\nn
\eea
Relabelling $(e^5,e^6,e^7)\rightarrow(e^1,e^2,e^3)$, from section 3,
we find that an $AdS$ region in QK or CLAG geometry admits three local $SU(3)$
structures; in terms of the $SU(2)$ forms $K^A$ on $\mathcal{N}_4$,
these are given by
\bea
e^A&=&e^7,\\
J^A&=&K^A+\frac{1}{2}\e^{ABC}e^{B}\w e^C,\\
\mbox{Re}\Omega^1&=&K^3\w e^2+K^2\w e^3,\\
\mbox{Im}\Omega^1&=&K^3\w e^3-K^2\w e^2,
\eea
together with permutations of the last two equations. In QK geometry,
each of these local structures must individually satisfy the
conditions for a local $AdS_3$ $SU(3)$ structure in K\"{a}hler-4
geometry. In CLAG geometry, the structure forms $e^3$, $J^3$ and
$\Omega^3$ must together satisfy the $AdS$ conditions in
K\"{a}hler-4 geometry, while the $e^A$, $J^A$, $\Omega^A$, $A=1,2$,
must satisfy the $AdS$ conditions in SLAG geometry. In each case, reducing
these conditions to a minimal necessary and sufficient independent
set, we get the results quoted in section 2.

\section{Explicit solutions}
In this section we show that the known solutions with $AdS_3$ factors
fit nicely in the framework developed in the previous sections. To do
so we consider solutions describing the near-horizon limit of
M5-branes wrapping SLAG 4-cylces and K\"{a}hler 4-cycles in a
$CY_4$. These were first found in \cite{kim} in seven-dimensional
gauged supergravity. Throughout this section we will follow the
notation and conventions of \cite{wrap}. 

For the known solutions, the eleven-dimensional metric can be put into the form
\begin{eqnarray}
m^2\dd s^2&=&\Delta^{-2/5}\left[\frac{a_1}{u^2}~\dd s^2(\mbb^{1,5-d})+a_2 \dd s^2(\Sigma_d)\right]\nonumber\\
	&\;&+\Delta^{4/5}\left[e^{2q\Lambda}u^{2c_1}\DD X^a\DD X^a+e^{-2p\Lambda}u^{2c_2}\dd X^\a\dd X^\a\right]\;, 
\end{eqnarray}
with $a=1\ldots p$ and $\alpha=1\ldots q$ and $p+q=5$. The constants $a_1$ and $a_2$ specify the relative size of the AdS factor and the $d$-cycle $\Sigma_d$. For the cases corresponding to M5-branes wrapping 4-cycles, we have $d=4$,  $p=4$ and $q=1$. The relevant values of the remaining constants  for the two cases that we will be discussing are summarized in Table \ref{constants}.
\begin{table}
\begin{center}
\begin{tabular}{l|c|c|c|c|c|c|c}
\hline\hline
&$p$&$q$&$a_1$&$a_2$&$\ee^{10\Lambda}$&$c_1$&$c_2$\\*[2pt]
\hline
SLAG 4-cycle in $CY_4$
   & 4 & 1 &
   $\ee^{4\Lambda}$ & $\ee^{-6\Lambda}$
   & $\frac{3}{2}$ & 1 & $\frac{3}{2}$ \\*[2pt]
K\"ahler 4-cycle in $CY_4$
   & 4 & 1 &
   $\ee^{4\Lambda}$ & $\ee^{-6\Lambda}$
   & $\frac{4}{3}$ & 1 & $\frac{4}{3}$ \\*[2pt]
\hline\hline
\end{tabular}
\end{center}
\caption{Known solutions of wrapped M5-branes on 4-cycles in a $CY_4$.}
\label{constants}
\end{table}
Following \cite{wrap}, we have defined
\begin{equation}
\DD X^a=\dd X^a+B^a_{\phantom a b}X^b
\end{equation}
where $B^a_{\phantom a b}$ is determined by the spin connection on $\Sigma_d$. Comparing this form of the metric with \eqref{1.4}, we identify
\begin{equation}
L=\frac{\Delta^{2/5}u^2}{a_1}\;,~~~~ C=\frac{\Delta^{2/5}u^{c_2}}{a_1}\;,
\end{equation}
where $e^9=C\dd z$ for the case at hand. Introducing new coordinates  
\begin{equation}
X^a=u^{-c_1}\cos\tau Y^a\;,~~~~ X^\a=u^{-c_2}\sin\tau Y^\a\;,~~~~c_1=e^{-2q\Lambda}\sqrt{a_1}\;,
~~~~c_2=e^{2p\Lambda}\sqrt{a_2}\;,
\end{equation}
where $Y^a$ and $Y^\a$ parametrise an $(p-1)$-sphere and an
$(q-1)$-sphere respectively, the metric becomes\footnote{This form of the metric corrects some errors in eq. 9.6 of \cite{wrap}.}
\begin{eqnarray}
m^2\dd s^2&=&\Delta^{-2/5}\left\{\frac{a_1}{u^2}~\left[\dd s^2(\mbb^{1,5-d})+\dd u^2\right]+a_2 \dd s^2(\Sigma_d)+e^{2(q-p)\Lambda}\dd\tau^2\right\}\nonumber\\
	&\;&+\Delta^{4/5}\left[e^{2q\Lambda}\cos\tau^2\DD Y^a\DD Y^a
	+e^{-2p\Lambda}\sin\tau^2\dd Y^\a\dd Y^\a\right]\;, 
\end{eqnarray}
From the definition of $\Delta$ in \cite{kim}, we have that
\bea
(a_1\l)^{-3}=\Delta^{-6/5}=e^{-2q\Lambda}\cos\tau+e^{2p\Lambda}\sin\tau.
\eea
Specialising to our cases of interest $p=4$ and $q=1$ from now on, the metric can be put into the generic form \eqref{b} by introducing a new coordinate $\rho$ as follows. Define
\begin{equation}
f(\rho)=a_1 c_2 e^{-p\Lambda}\sin\tau\;,
\end{equation}
where $f(\rho)$ is the same function as in section \ref{sec:coordsAdS}. Then, the metric becomes
\begin{eqnarray}
\dd s^2&=&\frac{1}{\lambda m^2}\Big[\dd s^2(AdS_3)+\frac{\lambda^3}{4(1-\l^3f^2)}\dd\r\otimes\dd\r+\frac{a_2}{a_1}\dd s^2(\Sigma_4)\nn&&+\frac{1}{c_1^2}(1-\lambda^3 f^2)\DD Y^a\otimes\DD Y^a\Big]\;,
\end{eqnarray}
where, in order to get the form \eqref{b}, we have set
\begin{equation}
f(\rho)=\frac{c_2}{2}\rho\;.
\end{equation}
From the analysis in section \ref{sec:coordsAdS}, we conclude that this solution carries electric flux, which is indeed the case for the solutions presented in \cite{kim}. 

To identify the $AdS$ radial coordinate in $\mathcal M_8$, one defines a one-form \cite{wrap}
\begin{eqnarray}
e^8&=&\frac{\Delta^{2/5}e^{q\Lambda}}{m}\left(c_1\cos\tau \frac{\dd u}{u}+\sin\tau \dd\tau\right)\nonumber\\
&=&\l^{-1/2}\sqrt{1-\l^3f^2}\dd r+\frac{\l^{5/2}f}{2m\sqrt{1-\l^3f^2}}\dd \r\;,
\end{eqnarray}
with $u=e^{mr}$. This expression matches our previous one in \eqref{eqn:e8}.

Now we are ready to check that the solutions of \cite{kim} satisfy our equations. We discuss the SLAG-4 and the K\"{a}hler-4 cases separately.

\subsubsection*{SLAG-4}  
For the SLAG-4 case, $B^a_{\phantom a b}=\bar\omega^a_{\phantom a b}$, where $\bar\omega^a_{\phantom a b}$ is the spin connection on $\Sigma_4$, which is just the four-hyperboloid $\mathcal{H}^4$ of unit curvature. As described in \cite{wrap}, the $SU(4)$ structure is given by
\begin{eqnarray}
J&=&e^a\w f^a\;,\\
\Omega&=&\frac{1}{4!}~\epsilon^{abcd}(e^a+if^a)(e^b+if^b)(e^c+if^c)(e^d+if^d)\;,
\end{eqnarray}
where $e^a=\Delta^{-1/5}\sqrt{a_2} m^{-1}\bar e^a$ and $f^a=\Delta^{2/5}e^{q\Lambda}u^{c_1}m^{-1}\DD X^a$. Here $\{\bar e^a\}$ denote a basis of $1$-forms on $\mathcal{H}^4$.  In the $AdS$ limit, the $SU(4)$ structure decomposes under $SU(3)$ in the following way:
\begin{eqnarray}
J&=&e^a\w(\tilde f^a-Y^ae^8)\nonumber\\
&=&J_{SU(3)}+e^7\w e^8\;,
\end{eqnarray}
where
\begin{equation}
e^7=-Y^a e^a\;,~~~~ \textrm{and}~~~~ 
\tilde f^a\equiv \frac{\Delta^{2/5}e^{q\Lambda}}{m}\cos\tau \DD Y^a\;.
\end{equation}
Similarly, the holomorphic 4-form $\Omega$ decomposes as
\begin{equation}
\Omega=(\mbox{Re}\Omega_{SU(3)}+i\mbox{Im}\Omega_{SU(3)})\w(e^7+ie^8)\;,
\end{equation}
from which we find
\begin{eqnarray}
\mbox{Re}\Omega_{SU(3)}=\frac{1}{3!}~\epsilon^{abcd} Y^a e^{bcd}
	-\frac{1}{2!}~\epsilon^{abcd} Y^a\tilde f^{bc}\w e^d\;,\\
\mbox{Im}\Omega_{SU(3)}=-\frac{1}{3!}~\epsilon^{abcd} Y^a \tilde f^{bcd}
	+\frac{1}{2!}~\epsilon^{abcd} Y^a\tilde f^b\w e^{cd}\;.
\end{eqnarray}

Using these expressions, it is straightforward to show that the SLAG-4 solution of \cite{kim} satisfies  \eqref{eqn:slag4Re}-\eqref{eqn:slag4de7}. Showing that \eqref{eqn:slag4Im} also holds requires some more work. Consider first the LHS of \eqref{eqn:slag4Im}. One should first note that
\begin{equation}
\DD^2 Y^a=\frac{1}{2}~\bar R^a_{\phantom a bcd}Y^b \bar e^c\w \bar e^d\;, 
\end{equation}
where $\bar R^a_{\phantom a bcd}$ is the Riemann tensor on $\mathcal{H}^4$ and hence $\bar R_{abcd}=2k\d_{a[c}\d_{d]b}$, with $k=-\frac{1}{3}$.  Then, we compute
\begin{eqnarray}
\tilde\dd\mbox{Im}\Omega=\frac{m\Delta^{-2/5}e^{-q\Lambda}}{\cos\tau}~\frac{1}{2!}\epsilon^{abcd}\left[\tilde
  f^{ab}e^{cd}+k\frac{\Delta^{6/5}e^{2q\Lambda}}{a_2}~\cos^2\tau~ Y^ae^b\left(\tilde f^{cd}-e^{cd}\right)e^7\right]
\end{eqnarray}
with $q=1$ for SLAG-4 \cite{wrap}. Taking the wedge product of the expression above with $\mbox{Im}\Omega$ we obtain
\begin{equation}
\mbox{Im}\Omega\w \dd\mbox{Im}\Omega= 6m\left(\frac{\Delta^{-2/5}e^{-q\Lambda}}{\cos\tau}+k\frac{\Delta^{4/5}e^{q\Lambda}}{a_2}~\cos\tau\right)~\mbox{Vol}_6\w e^7\;, \label{eqn:ImwdIm}
\end{equation}
where on a manifold with $SU(3)$ structure, the volume $6$-form may be defined by
\begin{eqnarray}
\mbox{Vol}_6&=&\frac{1}{3!}J\w J\w J\nonumber\\
	&=&-(e^{123}\tilde f^{123}+e^{124}\tilde f^{124}+e^{134}\tilde f^{134}+e^{234}\tilde f^{234})\;.
\end{eqnarray}

To put the result above into a more familiar form, we perform the following  change of coordinates: 
\begin{eqnarray}
\lambda=\frac{\Delta^{2/5}}{a_1}\;,\hspace{1cm} \sin\tau=\frac{e^{p\Lambda}}{a_1 c_2}~f(\r)\;,\hspace{1cm}
	\cos\tau=\frac{e^{q\Lambda}}{(a_1 \lambda)^{3/2}}~(1-\lambda^3f^2)^{1/2}\;.
\end{eqnarray}
Finally, introducing the values of $a_1$, $a_2$, $c_1$, $c_2$, for the SLAG-4 solutions of \cite{kim}, and using that $\cos\t=\l^{3/2}f(\r)$, \eqref{eqn:ImwdIm} can be cast in the following form:
\begin{equation}
\mbox{Im}\Omega\w \dd\mbox{Im}\Omega=\frac{3m\l^{1/2}}{\sin\t}\left(1+\cos^2\t\right)~\mbox{Vol}_6\w e^7\;.
\end{equation}
One can then show that this matches the RHS of
\eqref{eqn:slag4Im}. Furthermore, we have also checked that the
Bianchi identity for the four-form field strength is satisfied. To do
this, the following identities are useful:
\bea
\dd\left[\e^{abcd}Y^a\DD Y^b\w\DD Y^c\w\DD
  Y^d\right]&=&-\e^{abcd}Y^a\w\DD Y^b\w \DD Y^c\w
e^d\w(Y^ee^e),\nn
\dd\left[\e^{abcd}Y^a\DD Y^b\w e^c\w e^d\right]&=&-\left[2\e^{abcd}Y^a\DD
Y^b\w \DD Y^c\w e^d+\frac{1}{3}\e^{abcd}Y^ae^b\w e^c\w e^d\right]\nn&&\w
Y^ee^e,\nn
\dd[\e^{abcd}Y^ae^b\w e^c\w e^d]&=&-3\e^{abcd}Y^a\DD Y^b\w e^c\w
e^d\w (Y^ee^e),\nn
\e^{abcd}\DD Y^a\w\DD Y^b\w e^c\w e^d&=&-2\e^{abcd}Y^a\DD Y^b\w\DD
Y^c\w e^d\w(Y^ee^e).
\eea

\subsubsection*{K\"{a}hler-4}
For the K\"{a}hler-4 solutions presented in \cite{kim}, one can take $B^{12}=B^{34}$ with all other components vanishing. Then one has
\begin{equation}
B^{12}+B^{34}=-\frac{1}{2}~\bar\omega_{ab}K^{3ab}\;,
\end{equation}
where $K^3$ is defined in \eqref{eqn:ks}. Making the following ansatz for the $SU(4)$ structure,
\begin{eqnarray}
J&=&e^1\w e^2+e^3\w e^4+f^1\w f^2+f^3\w f^4\;,\\
\Omega&=&(e^1+ie^2)(e^3+ie^4)(f^1+if^2)(f^3+if^4)\;,
\end{eqnarray}
we find, in the $AdS$ limit, that the $SU(3)$ structure is given by
\begin{eqnarray}
e^7&=&K^3_{ab} Y^a\tilde f^b\;, \nonumber\\
J_{SU(3)}&=&e^1\w e^2+e^3\w e^4+\tilde f^1\w \tilde f^2+\tilde f^3\w \tilde f^4\;, \nonumber\\
\mbox{Re}\Omega_{SU(3)}&=&-K^2\w K^1_{ab} Y^a \tilde f^b-K^1\w K^2_{ab}Y^a\tilde f^b\;,\nonumber\\
\mbox{Im}\Omega_{SU(3)}&=&-K^2\w K^2_{ab}Y^a \tilde f^b+K^1\w K^1_{ab} Y^a\tilde f^b\;,
\end{eqnarray}
where the $K^A$, $A=1,2,3$, have been defined in \eqref{eqn:ks}. These expressions are just the same as those in eq.\eqref{eqn:K4su3}. Now it is a straightforward exercise to check that our equations \eqref{eqn:k4jj}-\eqref{eqn:k4de7} are satisfied and so is the Bianchi identity.

\section{Conclusions}
In this paper, we have formalised a proposal for a universal feature of the
global geometry of supergravity solutions of relevance to
the supersymmetric $AdS_3/CFT_2$ correspondence in M-theory. A supergravity solution associated to a CFT - a region of spacetime
containing a local $AdS$ region - should admit a
globally-defined $\mathbb{R}^{1,1}$ frame, and a global reduction of
its frame bundle, to one with structure group contained in $Spin(7)$. From
this starting 
assumption, we have seen how many individual features of AdS/CFT
geometry may be assembled into a coherent overall picture. Probe brane
kappa-symmetry projections arise from the global definition of the
spinorial realisation of the frame bundle. Solutions with
asymptotically vanishing flux
automatically asymptote to special holonomy manifolds. The existence
of a globally-defined frame bundle allows for the global truncation of
the field equations of eleven-dimensional supergravity. The
general necessary and sufficient conditions for minimally supersymmetric $AdS_3$
geometry in M-theory may be derived by imposing an $AdS_3$ 
boundary condition on the truncation of supergravity to a Cayley
frame bundle. The same applies for $AdS_2$ with an $SU(5)$
frame bundle \cite{eoin}, $AdS_4$ (with
magnetic fluxes) and a $G_2$ frame bundle; and $AdS_5$ with an $SU(3)$ frame bundle
\cite{wrap}. The minimal truncations, and associated $AdS$ conditions,
may be refined by further reducing the structure group of the frame
bundle and/or by demanding additional Killing spinor
realisations. Freund-Rubin or gauged supergravity $AdS$ solutions
satisfy the general equations for $AdS$ horizons in the appropriate
geometries.  

One of the original motivations for this work, and that of
\cite{wrap}, \cite{eoin}, was to map out the supersymmetric $AdS$
landscape of M-theory. At this point, it worth summarising what has
been achieved. The strategy in each of these papers is first to impose
the existence of a global Minkowski frame bundle, realised by Killing spinors of a
definite Minkowski chirality, and then to impose a general $AdS$
boundary condition on the global truncation of eleven-dimensional
supergravity to the Minkowski frame bundle. Modulo quotients, this
approach covers all supersymmetric $AdS$ spacetimes which may be
obtained from solutions with globally defined Minkowski frame
bundles, and definite chirality Minkowski
Killing spinors. For $AdS_2$, with the exception of near-horizon
limits of M5 branes wrapped on the direct product of a SLAG-3 and a
K\"{a}hler-2 cycle in a manifold of $SU(3)\times SU(2)$ holonomy, and
modulo some technical caveats, the results of \cite{eoin} are
complete. For $AdS_3$ with less than
sixteen supersymmetries, we believe that the combined results of this paper and
\cite{wrap} are
complete. We have certainly covered all cases which admit a wrapped brane
interpretation, and in full generality. There exist half-BPS $AdS_3$
solutions of M-theory; we have not performed a general investigation of this
interesting case here and we leave it for the future. For $AdS_4$
with electric fluxes, the Freund-Rubin solutions are exhaustive. For
$AdS_4$ with purely magnetic fluxes, admitting a wrapped brane
interpretation and modulo some technical caveats explained therein,
the results of \cite{38} and \cite{wrap} are complete. The existence or otherwise of
supersymmetric $AdS_4$ solutions with dyonic fluxes is an open
problem. For $AdS_5$ spacetimes admitting a wrapped brane
interpretation, and again modulo some technical caveats detailed in
\cite{wrap}, the results of \cite{39} and \cite{41} are
complete. This, then, is the status of the classification; the most
interesting cases that have not been covered are half-BPS $AdS_3$ and
dyonic $AdS_4$. If there exist any other $AdS$ solutions of M-theory
which are not covered by the classification, they will necessarily have very
complicated and unusual geometry.  

This global framework, and the results of the classification, open the
way for many future applications. The most obvious is to use the
geometrical insight provided by the $AdS$
torsion conditions to construct new explicit $AdS$ solutions. A more
important question is the development of a
theory of boundary conditions for solutions of the truncated
supergravity equations. With our general $AdS_3$ boundary condition,
we have taken a first step in this direction,
but there are many other possibilities to be explored. As we mentioned
before, a class of boundary conditions which is particularly
interesting mathematically is special holonomy spacelike asymptotics with
vanishing fluxes, and spacelike $AdS$ asymptopics associated to event
horizons. To our knowledge, the only known solutions of this form are
the elementary brane solutions, associated to interpolations from
conical special holonomy manifolds to Freund-Rubin $AdS$
horizons. These interpolating solutions are intimately
associated to the resolution of singularities of the asymptotic special
holonomy manifolds; and in these cases, the AdS/CFT correspondence 
provides a definition of how the singularities may be
resolved quantum gravitationally. This has been made manifest in the
work on the $Y^{p,q}$/quiver gauge theory correspondence in IIB. Open
string theory on the Calabi-Yau - with Dirichlet boundary conditions
restricting the open strings to the vicinity of the singularity -
reduces at low energies to a conformally invariant quiver gauge
theory. The AdS/CFT correspondence states that this field theory is
dual to the low-energy limit of closed IIB strings on a large-volume
$AdS_5\times Y^{p,q}$ manifold. The field theory at weak 't Hooft
coupling encodes the toric data of the Calabi-Yau singularity. It also
encodes, at strong 't Hooft coupling, the 
Sasaki-Einstein data of the $AdS$ manifold. This means that, at low
energies, the physical content of open string theory near a conical Calabi-Yau
singularity and closed string theory on a large-volume $AdS$ blow-up
of the singularity are contained in the same quantum field
theory. Going from weak to strong 't Hooft coupling in the field
theory gives a quantum definition of the singularity-resolving
geometrical interpolation. It would be very interesting if other
interpolating solutions, associated to the resolution of other types
of special holonomy 
singularities, could be constructed. Extending the intuition obtained
from conical interpolations, it seems likely that such solutions will
be associated to the resolution of singularities of collapsing
supersymmetric cycles of the asymptotic special holonomy
manifolds. Understanding how to construct such 
solutions - and how to associate a given $AdS$ near-horizon geometry
to a special holonomy infinity, and vice versa - will require a
detailed understanding of how to match the boundary data at each spacelike
infinity. It will certainly require the use of more sophisticated
geometrical techniques than those employed here. A complementary
approach to finding explicit interpolating solutions would be to try 
to establish existence or obstruction theorems. An example of such a
result - an obstruction theorem for Sasaki-Einstein metrics - has
recently appeared in the context of conical interpolations in IIB
\cite{yau}. It will be interesting to explore a more general extension
of this kind of approach. 

Another extension will be to consider more general
boundary conditions for solutions of the global supergravity
truncations - solutions with an $AdS$ region which asymptote to flux
geometries, or solutions with multiple $AdS$ regions. Known solutions
are likely to provide useful insights into the general form of boundary
conditions one should impose in these cases.

Though we have focussed on its classical, geometrical limit throughout
this paper, the AdS/CFT correspondence is of course a quantum
phenomenon. We have a much less satisfactory understanding of the
quantum aspects of AdS/CFT in M-theory than we do in IIB, and
improving this situation is a major outstanding 
problem. Perhaps the most promising line of attack is to exploit the duality
between M-theory and IIB, where things are much better
understood. This can work in two directions. By imposing $T^2$
isometries on the eleven-dimensional supergravity truncations - or their $AdS$
limits - one could obtain the subset of M-theory
solutions which can be reduced and T-dualised to IIB. This is how the
$Y^{p,q}$ were discovered. Conversely, where a IIB AdS/CFT dual
is known, if the geometry admits a lift to M-theory one could do so, in the
hope of gaining an understanding of how the known quantum field theory
encodes the eleven-dimensional geometrical data. Ultimately it might
be possible to extend the intuition thus obtained to M-theory
geometries not admitting a IIB reduction, though this will certainly
require significant new insights.

\section{Acknowledgements}
We are grateful to Bo Feng, Anastasios Petkou, Daniel Waldram
and in particular Jerome Gauntlett for useful discussions. PF was
supported by a BE and an FI 
fellowship from AGAUR (Generalitat de Catalunya), DURSI 2005 SGR
00082, CICYT FPA 2004-04582-C02-02 and EC FP6 program
MRTN-CT-2004-005104. OC is supported by EPSRC, and is grateful to the
University of Crete for hospitality during the final stages of this work.

\end{document}